\documentclass[preprint]{aastex}

\received{}
\accepted{}
\newcommand{\AAA}{$\rm{\AA} \;$}

\slugcomment{Accepted by AJ}

\author{
Timothy A. Reichard,\altaffilmark{1}
Gordon T. Richards,\altaffilmark{2}
Patrick B. Hall,\altaffilmark{2,3}
Donald P. Schneider,\altaffilmark{4}
Daniel E. Vanden Berk,\altaffilmark{5}
Xiaohui Fan,\altaffilmark{6}
Donald G. York,\altaffilmark{7,8}
G.R. Knapp,\altaffilmark{2}
and J. Brinkmann\altaffilmark{9}
}

\altaffiltext{1}{Department of Physics and Astronomy, The Johns Hopkins University, 3400 North Charles Street, Baltimore, MD 21218-2686.}
\altaffiltext{2}{Princeton University Observatory, Peyton Hall, Princeton, NJ 08544.}
\altaffiltext{3}{Departamento de Astronom\'{\i}a y Astrof\'{\i}sica, Pontificia Universidad Cat\'{o}lica de Chile, Casilla 306, Santiago 22, Chile.}
\altaffiltext{4}{Department of Astronomy and Astrophysics, The Pennsylvania State University, 525 Davey Laboratory, University Park, PA 16802.}
\altaffiltext{5}{Department of Physics and Astronomy, University of Pittsburgh, 3941 O'Hara Street, Pittsburgh, PA 15260.}
\altaffiltext{6}{Steward Observatory, University of Arizona, 933 North Cherry Avenue, Tucson, AZ 85721.}
\altaffiltext{7}{Department of Astronomy and Astrophysics, The University of Chicago, 5640 South Ellis Avenue, Chicago, IL 60637.}
\altaffiltext{8}{Enrico Fermi Institute, The University of Chicago, 5640 South Ellis Avenue, Chicago, IL 60637.}
\altaffiltext{9}{Apache Point Observatory, P.O. Box 59, Sunspot, NM 88349.}

\begin{document}

\title{Continuum and Emission-Line Properties of Broad Absorption Line Quasars}

\begin{abstract}

We investigate the continuum and emission-line properties of 224 broad
absorption line quasars (BALQSOs) with $0.9 \lesssim z \lesssim 4.4$
drawn from the Sloan Digital Sky Survey (SDSS) Early Data Release
(EDR), which contains 3814 bona fide quasars.  We find that
low-ionization BALQSOs (LoBALs) are significantly reddened as compared
to normal quasars, in agreement with previous work.  High-ionization
BALQSOs (HiBALs) are also more reddened than the average nonBALQSO.
Assuming SMC-like dust reddening at the quasar redshift, the amount of
reddening needed to explain HiBALs is $E(B-V) \sim 0.023$ and LoBALs
is $E(B-V) \sim 0.077$ (compared to the ensemble average of the entire
quasar sample).  We find that there are differences in the
emission-line properties between the average HiBAL, LoBAL, and nonBAL
quasar.  These differences, along with differences in the absorption
line troughs, may be related to intrinsic quasar properties such as
the slope of the intrinsic (unreddened) continuum; more extreme
absorption properties are correlated with bluer intrinsic continua.
Despite the differences among BALQSO sub-types and nonBALQSOs, BALQSOs
appear to be drawn from the same parent population as nonBALQSOs when
both are selected by their UV/optical properties.  We find that the
overall fraction of traditionally defined BALQSOs, after correcting
for color-dependent selection effects due to different SEDs of BALQSO
and nonBALQSOs, is $13.4\pm1.2$\% and shows no significant redshift
dependence for $1.7 \le z \le 3.45$.  After a rough completeness
correction for the effects of dust extinction, we find that
approximately one in every six quasars is a BALQSO.
\end{abstract}


\section{Introduction}

The continuum and emission-line properties of broad absorption line
quasars (BALQSOs) have been the subject of a number of previous
studies.  The primary goal of such work is to determine if BALQSOs are
drawn from the same parent population as nonBALQSOs.  If the average
continuum and emission-line properties of BALQSOs and nonBALQSOs are
in good agreement, then we can be reasonably certain that they are
drawn from the same parent population \markcite{wmf+91}({Weymann} {et~al.} 1991).  If there are
significant differences, it could be argued that BALQSOs are a
completely different class of quasar \markcite{sh87,bm92}({Surdej} \& {Hutsemekers} 1987; {Boroson} \& {Meyers} 1992).  It could also
be that the physical mechanism(s) that produce broad absorption line
(BAL) troughs are correlated with the continuum and emission-line
mechanisms.  For example, quasars seen at larger inclination angles,
with larger black hole masses, or with higher accretion rates may have
distinct continuum and emission-line properties, and they may also be
more or less likely to exhibit BAL troughs.

In the first such study, \markcite{wmf+91}{Weymann} {et~al.} (1991) combined 34 high-ionization
BALQSOs (HiBALs), 6 low-ionization BALQSOs (LoBALs) and 42 nonBALs to
make composite HiBAL, LoBAL and nonBAL spectra.  They used these
composites to show that the emission-line properties of BALs and
nonBALs are quite similar, except that the continua of LoBALs are
significantly redder than the continua of HiBALs and nonBALs.
\markcite{sf92}{Sprayberry} \& {Foltz} (1992) quantified this difference by determining that the
reddening of the LoBAL composite compared to the HiBAL composite could
be accounted for by dust reddening with a Small Magellanic Cloud (SMC;
\markcite{plp+84,pei92}{Prevot} {et~al.} 1984; {Pei} 1992) reddening law with $E(B - V) = 0.1$.

More recently \markcite{btb+01}{Brotherton} {et~al.} (2001) created composite BALQSO spectra from the
FIRST Bright Quasar Survey (FBQS; \markcite{wbg+00}{White} {et~al.} 2000).  Using a sample
of 25 HiBALs and 18 LoBALs, they also found that BALQSOs are redder
than the typical nonBALQSO and that the HiBAL and LoBAL spectra can be
brought into agreement with a composite nonBALQSO spectrum by applying
an SMC-type reddening law with $E(B-V)\sim0.04$ and $E(B-V)\sim0.1$,
respectively.  Note that the \markcite{btb+01}{Brotherton} {et~al.} (2001) criteria for BAL
classification is less strict than is traditional and several of their
BALQSOs have ``balnicity'' indices\footnote{As defined by
\markcite{wmf+91}{Weymann} {et~al.} (1991).} of zero or nearly zero and also that the FBQS sample
is selected at relatively red wavelengths, which should make the
sample more robust to loses from dust extinction.

In the past three years, a number of BALQSO studies utilizing the
Sloan Digital Sky Survey (SDSS; \markcite{yor+00}{York} {et~al.} 2000) have been published:
the Early Data Release (EDR) BALQSO catalog \markcite{rrh+02}({Reichard} {et~al.} 2003), the
analysis of the balnicity distribution and BALQSO fraction in
\markcite{kro+02}{Tolea}, {Krolik}, \& {Tsvetanov} (2002), the identification of a significant population of
unusual BALQSOs by \markcite{hal+02}{Hall} {et~al.} (2002), and the study of BAL-like intrinsic
absorption in a radio-detected sample of quasars by \markcite{mvi+01}{Menou} {et~al.} (2001).
In addition, radio \markcite{bwg+00,bwg+01}(e.g., {Becker} {et~al.} 2000, 2001), X-ray
\markcite{gbc+02}(e.g., {Gallagher} {et~al.} 2002) and rest-frame optical \markcite{bor02}(e.g., {Boroson} 2002)
observations of BALQSOs have provided important clues regarding the
nature of the BAL phenomenon.

In this paper, we add to these previous results by investigating
the continuum and emission-line properties of a catalog of 224 BALQSOs
\markcite{rrh+02}({Reichard} {et~al.} 2003) that were drawn from the SDSS EDR Quasar Catalog
\markcite{sch+02}({Schneider} {et~al.} 2002).  Using this BALQSO catalog, we create a number of
composite BALQSO spectra including: HiBAL (180 quasars), LoBAL (34
quasars) and FeLoBAL\footnote{FeLoBALs are low-ionization BALQSOs that
also have strong absorption from excited iron \markcite{bgh+97,hal+02}({Becker} {et~al.} 1997; {Hall} {et~al.} 2002).}
(10 quasars); in addition we constructed a nonBALQSO composite
spectrum from a set of 892 nonBALQSOs that have the same redshift and
apparent magnitude distribution as the objects in the BALQSO
catalog\footnote{Of course, if BALQSOs are extincted by dust, then our
BALQSOs will actually be more luminous than the (observed)
luminosity-matched sample of nonBALQSOs.}.  These spectra represent a
significant increase in the signal-to-noise ratio of composite BALQSO
spectra that can be used to investigate differences in continuum and
emission-line properties of BALQSOs and nonBALQSOs.  We examine these
differences from the perspective of whether the UV/optical properties
of BALQSOs and nonBALQSOs are consistent with being drawn from the
same parent sample (among optically-selected quasars).

The paper is structured as follows.  We review in \S~\ref{sec:data}
our initial quasar sample and the resulting BALQSO sample.
Section~\ref{sec:composite} discusses the construction and application
of our composite spectra.  We then analyze the broad-band color
(\S~\ref{sec:colors}) and UV/optical continuum
(\S~\ref{sec:continuum}) properties of BALQSOs.  An investigation of
the emission-line properties is given in in \S~\ref{sec:emline}.  In
\S~\ref{sec:parent} we comment on what our results mean in terms of
the BALQSO parent sample.  Using what we have learned about the colors
of BALQSOs we present an analysis of the BALQSO fraction in
\S~\ref{sec:balfractions}.  Finally, \S\ref{sec:conclusions}
summarizes our results.  Cosmology dependent parameters (e.g.,
absolute magnitude) are used in this paper only in a relative sense,
thus throughout this paper we will use a cosmology where $H_o
={\rm50\,km\,s^{-1}\,Mpc^{-1}}$, $\Omega_M = 1$, and $\Omega_\Lambda =
0$.  We also adopt a convention for optical spectral index such that
$\alpha=\alpha_{\lambda}$ unless stated otherwise, where $f_{\lambda}
\propto \lambda^{\alpha_{\lambda}}$.

\section{The Data \label{sec:data}}

Our analysis is based upon the 224 SDSS BALQSOs cataloged by
\markcite{rrh+02}{Reichard} {et~al.} (2003).  This catalog includes 185 BALQSOs from a complete
sample between $z=1.7$ and $z=4.2$, of which 131 quasars also form a
relatively homogeneous sample of objects that were selected as quasar
candidates using the SDSS's adopted quasar target selection algorithm
as described by \markcite{rfn+02}{Richards} {et~al.} (2002a).  This catalog was constructed from the
3814 bona fide quasars ($M_{i^*}\,<\,-23$, with at least one line
broader than $1000\,{\rm km\,s^{-1}}$) from the SDSS EDR quasar sample
\markcite{sch+02}({Schneider} {et~al.} 2002).  The EDR quasars were selected for spectroscopic
follow-up from the SDSS imaging survey, which uses a wide-field five
filter \markcite{fig+96}({Fukugita} {et~al.} 1996) multi-CCD camera \markcite{gcr+98}({Gunn} {et~al.} 1998) on a dedicated
2.5m telescope.  The spectra cover the optical range 3800$-$9200 \AAA
at a resolution 1800$-$2100.  The spectra are calibrated to
spectrophotometric standard F-stars and are thus corrected for
Galactic reddening (at the scale of the separation to the nearest
calibration star, $\lesssim1\deg$) in the limit that these F stars are
behind the dust in the Galaxy.  The spectrophotometric calibration is
thought to be good to $\sim$20\%.  Details of the photometric
calibrations are given by \markcite{hfs+01}{Hogg} {et~al.} (2001) and \markcite{stk+02}{Smith} {et~al.} (2002), and
details of the astrometric calibration are given by \markcite{pmh+03}{Pier} {et~al.} (2003).
The spectroscopic tiling algorithm is discussed by \markcite{blm+03}{Blanton} {et~al.} (2003).

The \markcite{rrh+02}{Reichard} {et~al.} (2003) EDR BALQSO catalog includes BALQSOs ($0.892 \le z
\le 4.41$) selected by two automated algorithms, cross-checked and
supplemented by a manual inspection.  One automated algorithm fitted a
composite quasar spectrum created from the entire EDR quasar
catalog\footnote{This is a composite spectrum of all 3814 quasars from
the EDR quasar catalog \markcite{sch+02}({Schneider} {et~al.} 2002). The composite was created as
described by \markcite{vrb+01}{Vanden Berk} {et~al.} (2001).} to each of the individual quasar spectra
in order to define the intrinsic flux level from which broad
absorption could be measured \markcite{rrh+02}({Reichard} {et~al.} 2003).  The algorithm also
scaled the emission line associated with the broad absorption (either
\ion{C}{4} or \ion{Mg}{2}) to create a more accurate continuum where
the broad-absorption and emission-line regions overlap.  The second,
more traditional algorithm, described by \markcite{kro+02}{Tolea} {et~al.} (2002), used a
power law to describe the underlying continuum and a Gaussian profile
to describe the emission-line region.  Since the blue half of the
emission line may be attenuated by absorption, a half-Gaussian was
fitted to the red half of the \ion{C}{4} emission line and the
half-Gaussian was replicated to the blue half of the \ion{C}{4}
emission line.  Both algorithms then computed the standard balnicity
index \markcite{wmf+91}({Weymann} {et~al.} 1991) of \ion{C}{4} absorption to classify objects as
HiBALs or as objects without detectable BAL troughs --- nonBALs.  The
EDR spectra were also inspected by eye (by P.~B.~H.)  and were
classified as nonBALs, HiBALs, LoBALs, or FeLoBALs.  Further details
regarding the construction of the sample can be found in the catalog
paper, \markcite{rrh+02}{Reichard} {et~al.} (2003).

\section{BALQSO Composite Construction and Fitting \label{sec:composite}}

\subsection{BALQSO Composite Spectra}

\markcite{rrh+02}{Reichard} {et~al.} (2003) created geometric mean composite spectra of the classes
of BALQSOs.  The 185 complete-sample BALQSOs and 39 supplementary
BALQSOs were partitioned into a HiBAL class (180 objects with
high-ionization broad absorption troughs, such as \ion{C}{4}, but
without low-ionization broad absorption lines, such as \ion{Mg}{2}), a
LoBAL class (34 objects with high- and low-ionization broad absorption
lines), a combined class of both HiBALs and LoBALs, and the small
class of FeLoBALs.  \markcite{rrh+02}{Reichard} {et~al.} (2003) also created a nonBAL sample and
composite spectrum of 892 objects with similar redshift and absolute
$i^*$ magnitudes to the total BALQSO sample.  The composite spectra
(including the EDR composite spectrum), normalized to unity at a rest
wavelength of 2500 \AA, are reproduced in Figure~\ref{fig:fig1};
the size and properties (see below) of the samples are given in
Table~1.  The signal-to-noise ratio is significantly lower in the
LoBAL and FeLoBAL composite spectra than in the other composites
because of the small sample sizes.

A composite spectrum created from a sample of reddened power-law
spectra will possess a mean spectral index and a mean reddening.  For
geometric mean composites, the composite spectral index and reddening
are the arithmetic mean spectral index and arithmetic mean reddening
of the individual spectra, as we show in the Appendix.  Each of our
composite spectra are formed by the geometric mean approach.  Thus
comparisons of spectral indices and reddening in different composite
spectra are equivalent to comparisons of the general subsamples that
comprise the composites.

\subsection{Continuum Fitting with a Composite Spectrum \label{sec:fitting}}

In our investigation of the continuum differences between quasar
types, we will rely upon the composite-fitting algorithm described by
\markcite{rrh+02}{Reichard} {et~al.} (2003), since that method allows some freedom in disentangling
the power-law continuum from any curvature in the spectrum (from dust
extinction, for example).  The method is explained in detail by
\markcite{rrh+02}{Reichard} {et~al.} (2003), but we provide a brief overview here.

Quasar continua are often described by a power law relation
$f_{\lambda} \propto \lambda^{\alpha}$ with a spectral index $\alpha
\; (= \alpha_{\lambda} = -2 - \alpha_{\nu})$ and roughly Gaussian
emission-line profiles.  In this paper we make the assumption that
deviations from a power law times the template composite spectrum over
a wide wavelength range can be modeled as extinction by dust.  The
continuum is represented by
\begin{equation}
f_{\lambda} \propto g_{\lambda} \lambda^{\alpha} 10^{-aE(B-V) \xi(\lambda)},
\end{equation}
where $\lambda$ is in units of microns, $g_{\lambda}$ is the template
composite spectrum, $a = 0.4(1+R_V)$, and $\xi(\lambda)$ is the
extinction curve.  We have adopted the SMC extinction curve fit given
by \markcite{pei92}{Pei} (1992), for which the ratio of total to selective absorption
is $R_V = A_V/E(B-V) = 2.93$.  We choose the SMC curve because ---
unlike those of the LMC and the Milky Way --- it lacks the
``2200\,\AA\ bump'', which has never been conclusively seen from
quasar host galaxy dust, possibly due to the destruction of small
graphite grains by the quasar's radiation field
\markcite{plf03}(e.g., {Perna}, {Lazzati}, \& {Fiore} 2003).

We investigate the continuum properties with three different fitting
procedures.  If spectral indices alone are the reason that BALQSOs are
intrinsically redder than nonBALQSOs, then the BALQSO and nonBALQSO
composite spectra might be matched by a simple change of spectral
index.  On the other hand, if BALQSOs are really dust reddened and/or
if they do not have the same underlying spectral index distribution as
nonBALQSOs, then the fits will need to properly account for the lost
continuum flux at bluer wavelengths due to dust reddening.  Thus, one
procedure is a two-parameter fit, where we fit spectra by changing
both the power-law spectral index and $E(B-V)$ of a template composite
spectrum to match each individual spectrum.  We will also fit spectra
by allowing only one of these parameters to change.  To distinguish
these different fits, we will abbreviate them as $P$, for change in
the power-law spectral index only, $f_\lambda \propto
\lambda^{\alpha}$; $R$, for change in reddening only, $f_\lambda
\propto 10^{-aE(B-V)\xi(\lambda)}$; and $RP$, for change in both the
power-law spectral index and reddening, $f_\lambda \propto
\lambda^{\alpha} 10^{-aE(B-V)\xi(\lambda)}$.

The template composite spectrum that we use for our fits is the EDR
composite spectrum (all 3814 objects).  Although we do not know how
much reddening is included in the EDR composite, we arbitrarily define
the EDR composite reddening to be zero.  Thus, the measured reddening
values do not indicate absolute reddening but rather reddening
relative to the EDR composite spectrum.  For reference, we computed
the spectral index of the EDR composite spectrum to be
$\alpha_{\lambda}=-1.58$ by fitting a power law to points near
$\lambda = $ 1355 \AAA and 2240 \AA.

\section{Broad-band UV/Optical Colors of BALQSOs \label{sec:colors}}

We can begin addressing the question of the continuum differences
between BALQSOs and nonBALQSOs by investigating the differences
between their broad-band colors.  \markcite{mvi+01}{Menou} {et~al.} (2001) and \markcite{kro+02}{Tolea} {et~al.} (2002)
have already shown that the broad-band colors of BALQSOs are redder
than those of nonBALQSOs.  In the color-color plots in
Figure~\ref{fig:fig6} we expand upon this result using the full sample
of \markcite{rrh+02}{Reichard} {et~al.} (2003).  The panels in Figure~\ref{fig:fig6} include all
EDR nonBALQSOs (gray ``x''s) and EDR BALQSOs (open black squares,
HiBALs; filled black triangles, LoBALs) with redshifts between 1.8 and
3.5 (where our BALQSO sample is most complete).  It is clear that
BALQSOs are redder than nonBALQSOs on the average.  In particular, it
is obvious that the apparently bluest nonBALQSOs (e.g.,
$g^*-r^*\lesssim0$) are greatly underrepresented among the BALQSOs.
This result is {\em not} simply because of absorption from the BAL
troughs themselves.  The trough absorption can make the broad-band
color of BALQSOs bluer as well as redder, depending on where the
redshift of the quasar places the troughs with respect to the filters.
Instead, it is the overall flux deficit which spans the entire width
of a filter that causes BALQSOs to be redder than nonBALQSOs.

In the bottom right-hand panel, we show that BALQSOs are redder than
nonBALQSOs in yet another way.  Here we use relative colors
\markcite{rfs+01,rhv+02}({Richards} {et~al.} 2001; {Richards et al.} 2003), which are colors corrected by the median color
as a function of redshift and thus are independent of redshift.  From
these distributions of relative $g^*-i^*$ color, we find that BALQSOs
have an even more pronounced tail of red colors than do nonBALQSOs.

\section{BALQSO/nonBALQSO Continuum Comparison \label{sec:continuum}}

Extending the broad-band results, using spectral analysis, we now
investigate the finding of previous authors that LoBALs are redder
than both nonBALs and HiBALs \markcite{wmf+91}(e.g., {Weymann} {et~al.} 1991) and that even
HiBALs are redder than nonBALs \markcite{btb+01}(e.g., {Brotherton} {et~al.} 2001).  We examine
the continuum (i.e., color) differences between our BALQSO and
nonBALQSO samples in two ways: first by comparing the ensemble
averages in the form of composite spectra and second in terms of the
spectral indices and reddening values that result from our fits of
each individual quasar to a template.  Emission and absorption
properties of BALQSOs will be discussed in \S~\ref{sec:emline}.

\subsection{Composite Spectra Measurements\label{sec:compred}}

To compare the composite spectra continua, the EDR composite spectrum
is fitted to each of the BALQSO composite spectra by the
$\chi^2$-minimization procedure described in \S~3 of \markcite{rrh+02}{Reichard} {et~al.} (2003).
In short, the spectra are compared in wavelength regions where
emission and absorption are typically absent.  The algorithm
iteratively searches for values of the spectral index and reddening
that provides the best fit.  The resulting $\alpha_P$,
[$\alpha_{RP},E(B-V)_{RP}$], and $E(B-V)_{R}$ values are given in
Table~1.  The single-parameter fits yield consistent results that
suggest increased reddening with increasing BAL strength.  The
two-parameter fits are more difficult to interpret, especially given
the strong degeneracy between spectral index and reddening when
performing a two-parameter fit; see Figure~3 in \markcite{rrh+02}{Reichard} {et~al.} (2003).

As a simplistic check on our method and code, we fit the EDR quasar
composite spectrum to itself and indeed found that the algorithm
returns the appropriate values, specifically, $\alpha_P = \alpha_{RP}
= -1.58$ and relative reddening $E_R = E_{RP} = 0$.  The nonBAL
composite is slightly bluer\footnote{More negative values of $\alpha$
indicate bluer spectra.} ($\alpha_P = -1.61$) than the EDR composite
spectrum ($\alpha_P = -1.58$) and is best fit by lower reddening
values than the EDR composite. This result is to be expected if
BALQSOs are redder than nonBALQSOs since the EDR composite spectrum
includes BALQSOs.

We find that both the HiBAL and HiBAL+LoBAL composite spectra are
redder than the nonBALQSO composite spectrum, having power-law fits
with spectral indices $\alpha_P = -1.39$ and $-1.29$, respectively.
In addition, for the $R$-fit model, both of these composites require
reddening relative to the EDR composite spectrum: $E(B-V)=0.023$ and
$0.032$, respectively.  The redder values for the HiBAL+LoBAL composite
spectrum suggest that the LoBAL population is redder than the HiBAL
population.  Indeed, the LoBAL composite spectrum has the reddest
power-law-only fit, $\alpha_P=-0.93$, and requires the largest
correction for the dust-only fit, $E_{R}(B-V)=0.077$.  A large
reddening value is also required for the $RP$-fit of the LoBAL
composite.  However, the interpretation of $RP$-fit values is less
clear, and we are reluctant to place any absolute physical importance
on these values due to the degeneracy described above (but see the
next section, as well as \S~\ref{sec:parent} below).  Nevertheless,
the single-parameter fits unambiguously demonstrate that the sequence
of nonBALs, HiBALs, and LoBALs is a sequence of quasars with
increasingly red colors.  This result is consistent with the finding
of \markcite{btb+01}{Brotherton} {et~al.} (2001) for their radio-selected sample of HiBALs and
inconsistent with the original findings of \markcite{wmf+91}({Weymann} {et~al.} 1991); both of
these previous results are based on much smaller samples.

\subsection{Individual Measurements}

We derive qualitatively similar results from the fits to individual
quasars, as seen in Figure~\ref{fig:fig2}.  The upper, left-hand panel
shows that, for the $P$-fits, BALQSOs (both HiBALs and LoBALs) have a
distribution of spectral indices which is shifted slightly to the red
(more positive values of $\alpha_P=\alpha_{\lambda}$) and has a much
larger fraction of quasars in the red tail.  For the $RP$-fits (upper,
right-hand and lower, left-hand panels) the derived spectral indices
for both BALQSOs and nonBALQSOs are similar, but the best-fit
reddening values are skewed slightly to the red.

The $RP$-fit results are better illustrated in the lower, right-hand
panel where we plot $\alpha_{RP}$ versus $E(B-V)_{RP}$.  The
$\alpha_{RP}$ values are clearly non-physical since the range of
values is much larger than the observed distribution of spectral
indices in quasars and since there is a degeneracy between $E(B-V)$
and $\alpha$.  Although there is a true $\chi^2$-minimizing pair of
spectral index and reddening values for the $RP$-fits, the degeneracy
between $E(B-V)$ and $\alpha$ means that the two-parameter $RP$
fitting method cannot be used to determine unique, physical spectral
indices and reddening values.  As such, the $RP$ spectral indices and
reddening values must be considered only mathematical values to
reproduce the continua.  Nevertheless, we can clearly see that {\em at
a given value of $\alpha_{RP}$}, BALQSOs (LoBALs in particular) are
redder than nonBALQSOs as measured by $E(B-V)_{RP}$; this result
cannot be caused by the degeneracy between the two parameters of the
$RP$-fit.  A two-dimensional K-S test \markcite{ff87}({Fasano} \& {Franceschini} 1987) shows that the
nonBALQSO and BALQSO distributions in the [$E(B-V)_{RP},\alpha_{RP}$]
plane are different at the 99.9999\% confidence level
($\simeq5\sigma$).  The HiBAL and LoBAL (including FeLoBAL)
distributions are different at the 99.56\% confidence level
($\simeq2.8\sigma$).  Finally, as we shall discuss in
\S~\ref{sec:colortrends}, we have reason to believe that the $RP$-fit
values may still be useful in a relative sense.

\subsection{The Form of the Reddening\label{sec:red}}

Having shown that both LoBALs {\em and} HiBALs are redder than
nonBALQSOs and that LoBALs are redder than HiBALs, we explore the
question of the origin of the reddening.  Specifically, we ask whether
BALQSOs simply have redder power-law spectral indices or if they are
more dust reddened.  A combination of both is also possible, as are
non-dust-related reddening scenarios; for example Figures~12--14 in
\markcite{hab+00}{Hubeny} {et~al.} (2000) show that redder quasars might simply be observed more
face-on, have lower accretion rates, and/or have larger black hole
masses, respectively.  Since previous work has shown that SMC-like
dust reddening might be able to account for the redness of BALQSOs, we
adopt that law for our purposes here.

To aid in our analysis, we present the differences between the nonBAL,
HiBAL, and LoBAL composite spectra in
Figures~\ref{fig:fig3}--\ref{fig:fig5}.  In each case we present the
difference spectra that result from modifying the BALQSO composites by
all three of our fitting methods ($P$, $R$, and $RP$) in addition to
showing the unaltered difference spectra (i.e., no changes in spectral
index or reddening, but are normalized to unity at 2900 \AA).  Because the EDR and
nonBAL composite spectra are similar, as are the HiBAL+LoBAL and HiBAL
composite spectra, we have chosen to use only the composite spectra of
the mutually exclusive nonBAL, HiBAL, and LoBAL classes in the rest of
our analysis.

The $P$-fit (top spectrum in Figs.~\ref{fig:fig3}--\ref{fig:fig5}) was
achieved by changing only the spectral index $\alpha_P$ of each of the
HiBAL and LoBAL composites.  From the $P$-fits, one could make an
argument for BALQSOs simply having redder spectral indices than
nonBALQSOs.  Similarly, applying only the SMC ($R$-fit) reddening law
to the composites (second spectrum in
Figs.~\ref{fig:fig3}--\ref{fig:fig5}) suggests that dust-only reddening
might just as adequately account for the color differences between
BALQSOs and nonBALQSOs.  Finally, the $RP$-fit difference spectra are
plotted as the third spectrum in each of
Figures~\ref{fig:fig3}--\ref{fig:fig5}, using spectral indices and
relative reddening from Table~\ref{tab:tab1}.  As with both of the
previous fitting procedures, the $RP$-fit does a reasonably good job
of correcting for the color differences between the BALQSOs and the
nonBALs.

For wavelengths longer than about $1700\,{\rm \AA}$, each of the three
reddening corrections to the BALQSOs reproduces the continuum of the
nonBALs reasonably well.  Thus, it is difficult to use this analysis
alone to answer the question posed above regarding the nature of the
reddening in BALQSOs.  Expanding our wavelength baseline via UV and
near-IR spectra of our objects would clearly be valuable in terms of
answering the question of the form of the reddening.  Nevertheless, we
have shown that the reddening that is observed in our BALQSO sample is
certainly {\em consistent} with SMC-like dust as suggested by previous
authors.  Furthermore, at least some contribution from reddening is
favored for LoBALs, as seen by the $\chi^2$ of the fits in
Figures~\ref{fig:fig4} and~\ref{fig:fig5} and our analysis in
\S~\ref{sec:colortrends}.

If BALQSOs and nonBALQSOs have the same intrinsic spectral index
distribution and BALQSOs are reddened by SMC-like dust, then the
average reddening for HiBALs is $E(B-V)=0.023$ and for LoBALs is
$E(B-V)=0.077$ (as compared to the EDR composite spectrum, which is
reddened by $E(B-V)\sim0.004$ as compared to the nonBALQSO
composite).  These numbers can be compared to \markcite{sf92}{Sprayberry} \& {Foltz} (1992) and
\markcite{btb+01}{Brotherton} {et~al.} (2001) who each applied an SMC reddening law and found that
$E_{Lo} - E_{Hi} \sim 0.1$ was needed to account for flux deficits in
LoBALs as compared to HiBALs.  Our results suggest a smaller
differential reddening between the HiBALs and LoBALs.  This small
difference between our results and previous results may be related to
the fact that the SDSS EDR BALQSO catalog includes a much larger
fraction of BALQSOs with smaller balnicity indices than the
\markcite{wmf+91}{Weymann} {et~al.} (1991) sample.  One way that our smaller differential
reddening result might be understood is if BALQSOs with larger
balnicity indices are more heavily reddened; a larger sample is needed
to determine if this is actually the case.

\section{Emission and Absorption Properties \label{sec:emline}}

In addition to investigating the differences in the continua and
colors of BALQSOs, it is also important to examine the differences and
similarities in their emission lines.  In particular, if BALQSOs and
nonBALQSOs are drawn from the same parent population, they are likely
to have similar emission-line properties \markcite{wmf+91}({Weymann} {et~al.} 1991).  Thus, we now
revisit Figures~\ref{fig:fig3}--\ref{fig:fig5} with
emission lines instead of continua in mind.  At the same time we will
also comment on significant differences in the absorption line
properties of the different BALQSO samples.

We begin the emission-line analysis by examining the emission-line
region near \ion{Mg}{2}.  In each panel of
Figure~\ref{fig:fig3} the \ion{Mg}{2} emission lines are
similar for HiBALs and nonBALs, except that there appears to be a
slight dearth of emission in HiBALs just redward of the expected line
peak (as evidenced by negative flux in the HiBAL minus nonBAL
difference spectrum).  This result holds true as well for the LoBAL
composite in Figure~\ref{fig:fig4}, where we see evidence for
a broader decrement of flux in the red wing of \ion{Mg}{2}.  We also
note that strong, prominent absorption troughs, such as those found
blueward of \ion{C}{4} and \ion{Al}{3}, are absent blueward of
\ion{Mg}{2} in the LoBAL composite.  This absence is not unexpected
since the creation of composite spectra tends to wipe out features that
are not persistent from spectrum to spectrum.  Since the \ion{Mg}{2}
absorption troughs can occur at different velocities from the emission
line peak, the apparent strength of the absorption is diminished in
the composite spectrum.

Moving to shorter wavelengths, we find that the HiBAL composite shows
no excess in the UV48 complex of \ion{Fe}{3} at $2080\,{\rm \AA}$, but
that there may be a slight excess in LoBALs.  \markcite{wmf+91}{Weymann} {et~al.} (1991) reported
an excess at this wavelength among their LoBALs and also found that
the strength of this feature correlates with the balnicity index
for HiBALs as well as LoBALs.

Continuing on to the complex of emission near \ion{C}{3}]
$\lambda1909$, we find a small excess of emission in the \ion{Al}{3}
and \ion{Si}{3}] region of the HiBAL composite, which \markcite{wmf+91}{Weymann} {et~al.} (1991)
suggested is likely due to \ion{Fe}{3} or \ion{Fe}{2} emission.
However, we do not see any excess longward of the peak of \ion{C}{3}]
which would be expected from \ion{Fe}{3} UV34.  In the LoBAL
composite, this same excess may still persist, but more significant is
a decrement of flux at the centroid of \ion{C}{3}].

No strong absorption from \ion{Al}{3} is seen in the HiBAL composite,
but the LoBAL composite exhibits two prominent absorption troughs near
$1804\,{\rm \AA}$ and $1835\,{\rm \AA}$, with a peak near $1819\,{\rm
\AA}$.  A third depression is found near $1750\,{\rm \AA}$ and may
have a weaker counterpart in the HiBAL composite.  The \ion{Al}{3}
absorption lines are quite strong in the LoBAL composite, whereas the
\ion{Mg}{2} troughs which were used to define the LoBALs are
relatively absent, suggesting perhaps that \ion{Al}{3} is a better
indicator of LoBALs or that \ion{Al}{3} absorption has a more
consistent velocity distribution from quasar to quasar.

In the composite spectra shown in Figure~\ref{fig:fig1}, \ion{He}{2}
$\lambda1640$ emission appears to be slightly weaker in BALQSOs than
in nonBALQSOs.  The difference spectra in Figures~\ref{fig:fig3} and
\ref{fig:fig4} also exhibit this weakness.

At the location of \ion{C}{4} we see some interesting differences
in the spectra.  First, we confirm the reality of the ``droop'' in the
difference spectrum that was reported by \markcite{wmf+91}{Weymann} {et~al.} (1991).  This ``droop''
occurs between the \ion{C}{4} emission peak and $\sim1600\,{\rm \AA}$
and is seen in both the HiBAL composite and LoBAL composite, as shown
in Figures~\ref{fig:fig3} and~\ref{fig:fig4}.  We note
that the form of this ``droop'' is similar to the \ion{C}{4} emission
line flux decrement that was observed in quasars with large \ion{C}{4}
emission-line blueshifts \markcite{rvr+02}({Richards} {et~al.} 2002b) and may be an indication that
BALQSOs have larger-than-average \ion{C}{4} emission-line blueshifts;
see \S~\ref{sec:parent}.

In addition to this ``droop'' is the obvious decrement of flux
blueward of the emission-line peak that is caused by the \ion{C}{4}
BAL troughs themselves.  The absorption becomes significantly more
pronounced in the HiBAL composite near the expected peak of \ion{C}{4}
(i.e., at zero velocity with respect to the quasar redshift); in the
LoBAL composite, the change at zero velocity is even more dramatic.
\ion{C}{4} broad absorption troughs thus penetrate to velocities less
than the minimum velocity of the \ion{C}{4} balnicity index
($3000\,{\rm km\,s^{-1}}$ at 1534.5 \AA).  The \ion{C}{4} trough in
the HiBAL composite has a minimum near $1520\,{\rm \AA}$ with a weaker
shoulder that extends from about $1500\,{\rm \AA}$ to perhaps as blue
as the peak of \ion{Si}{4}+\ion{O}{4}].  The LoBAL composite
difference spectrum exhibits two distinct \ion{C}{4} troughs near
$1502\,{\rm \AA}$ and $1541\,{\rm \AA}$ that are separated by a peak
near $1518\,{\rm \AA}$ (which is roughly consistent with the 1520 \AAA
minimum of the \ion{C}{4} trough in the HiBAL composite difference
spectrum).  The structure in the composite spectra troughs is
particularly interesting since, as described above, the construction
of composites tends to average out any structure that is not
persistent from quasar to quasar.

There is absorption just blueward of the \ion{Si}{4}+\ion{O}{4}]
emission-line peak in both the HiBAL and LoBAL composite spectra.  In
the LoBAL composite difference spectrum, the absorption is
double-troughed (with minima near $1356\,{\rm \AA}$ and $1386\,{\rm
\AA}$ and separated by a peak near $1370\,{\rm \AA}$).  We see no
obvious trend in the separation of the double troughs in \ion{Al}{3},
\ion{C}{4} and \ion{Si}{4} that are seen in the LoBAL composite
difference spectrum, but for all three of these systems the peak
between the troughs is roughly $6000\,{\rm km\,s^{-1}}$ from the
expected emission-line center, which is suggestive of the radiative
acceleration ``ghost of Ly$\alpha$'' effect described by
\markcite{ara96}{Arav} (1996).

Toward the bluer end of the spectra, the continuum fits are not nearly
as good and it becomes difficult to tell whether the features that we
see are the result of excess emission or absorption.  Nevertheless,
there are a few notable features.  The LoBAL composite difference
spectrum displays absorption just blueward of each of the \ion{C}{2}
$\lambda 1335$, \ion{O}{1}/\ion{Si}{2} $\lambda 1304$ and \ion{Si}{2}
$\lambda 1262$ emission lines.  Finally, in the HiBAL composite
difference spectrum there is an excess of flux between \ion{N}{5} and
\ion{Si}{2} that may be similar to the excess \ion{N}{5} emission that
\markcite{wmf+91}{Weymann} {et~al.} (1991) reported for their BALQSOs.

\section{The BALQSO Parent Sample \label{sec:parent}}

\subsection{Comparison with Color and Emission-Line Shift Composites}

One of the primary reasons for comparing the continua and emission
lines of BALQSOs to nonBALQSOs is to determine if they are drawn from
the same parent population.  Although our BALQSOs and nonBALQSOs are
selected in the same manner using only the optical/UV part of the
spectrum, the SDSS selection algorithm is sensitive to a wide range of
optical/UV properties.  As such, we might expect to see differences in
the UV/optical properties of BALQSOs if they are indeed drawn from a
different parent sample.  Such differences should be more noticeable
at other wavelengths, particularly the IR and sub-mm (where dust is
seen in emission rather than absorption); however, recent sub-mm
observations show no significant differences between BALQSOs and
nonBALQSOs \markcite{lck03,wrg03}({Lewis et al.} 2003; {Willott et al.} 2003).

We consider the issue of the BALQSO parent sample by examining the
continuum and emission-line properties of BALQSOs as compared to
normal quasars.  In previous investigations of SDSS quasars, we found
significant and possibly related trends in the spectra as a function
of both the \ion{C}{4} emission-line blueshift \markcite{rvr+02}({Richards} {et~al.} 2002b) and the
broad-band colors \markcite{rhv+02}({Richards et al.} 2003).  In Figure 7, we plot the relative
$g-i$ color, $\Delta(g-i)$, against \ion{C}{4} blueshift for bright
($i<18.1$) EDR quasars.  We find a weak anti-correlation: quasars with
little or no blueshift are rarely as blue as $\Delta(g-i)\sim-0.1$,
while relative colors near $-0.1$ are typical for objects with high
blueshifts.  Quasars with larger \ion{C}{4} blueshifts thus appear to
have bluer optical colors.

Since previous work has shown that the emission features change as a
function of \ion{C}{4} blueshift and color, we suggest that it is
possible to learn something about the intrinsic properties of BALQSOs
by comparing their emission lines to the emission lines in the
\ion{C}{4} blueshift and color composites.  In
Figure~\ref{fig:fig8}, we compare our HiBAL and LoBAL composite
spectra in the spectral region covering the \ion{C}{4} and \ion{C}{3}]
emission lines to the \ion{C}{4} shift composites (left panels) from
\markcite{rvr+02}{Richards} {et~al.} (2002b) and to the color composite spectra (right panels) from
\markcite{rhv+02}{Richards et al.} (2003).  We specifically examine the composites in the red
wing of the \ion{C}{4} emission line and in the vicinity of the
\ion{He}{2}\,$1640\,{\rm \AA}$ and \ion{C}{3}] emission lines.  Since
we wish to concentrate on the similarities and differences of the
emission lines in these composites, we have first removed the overall
continua from each of the composites using the $RP$-fit method
described above.

In the left panels of Figure~\ref{fig:fig8}, we see that the small
\ion{C}{4} shift composite is a rather poor fit to either the HiBAL or
LoBAL composite spectra since the difference spectra deviate from zero
in the red wing of \ion{C}{4} emission, near \ion{He}{2} and at the
peak of \ion{C}{3}]).  On the other hand, the large \ion{C}{4} shift
composite appears to over correct for the ``droop'' redward of
\ion{C}{4} emission.  Both HiBALs and LoBALs appear to be best fit (in
the red wing of \ion{C}{4} and near both \ion{He}{2} and \ion{C}{3}])
by the composite with the second largest \ion{C}{4} emission-line
blueshift, which suggests that BALQSOs may be preferentially found
among quasars with large \ion{C}{4} shifts.  Since BALQSOs are
intermediate between the extremes of measured \ion{C}{4} blueshifts, a
distinct BALQSO parent population is not needed.

Similarly, when we compare our BALQSO composites to the composite
spectra as a function of color, we find that the BALQSOs are not
equally well described by each of the color composites.  The reddest
composite disagrees more in the red wing of \ion{C}{4} and at the peak
of \ion{C}{3}] than the bluest composite, which provides the best fit.
Thus BALQSOs may be drawn from a part of the parent sample which is
intrinsically bluer than average.  As with the blueshifts, although
there are differences in color between BALQSOs and nonBALQSOs, they
are not larger than the known differences among nonBALQSOs themselves,
which suggests that there is no need for a separate parent population.

\subsection{BALQSO Composites as a Function of Fitted Spectral Indices\label{sec:colortrends}}

The above results suggest that BALQSOs tend to have blue intrinsic
optical colors and moderately large \ion{C}{4} blueshift; however, the
situation is more complicated than that.  Another way to address the
issue of the intrinsic color of BALQSOs is to use the results from our
$RP$-fits to the individual quasar spectra.  That is, we can ask
whether the {\em relative} spectral indices that result from these
$RP$-fits are meaningful even though we know that the absolute values
are not necessarily physical.  If we assume that the relative
$\alpha_{RP}$ values are both meaningful and representative of
intrinsic color, we can create and compare composite HiBAL spectra as
a function of "intrinsic" color.  We have divided the ``complete''
sample of 124 HiBAL quasars from the EDR BALQSO catalog into three
equal parts according to their $\alpha_{RP}$ values.  The bluest and
reddest composites that result from these subsamples are shown in
Figure~\ref{fig:fig9}; the middle composite (not shown) has
intermediate properties.

We see that the red HiBALs have emission lines with much larger
equivalent widths --- consistent with the results of \markcite{rhv+02}{Richards et al.} (2003),
where bluer quasars were found to have weaker emission lines than
redder quasars.  There also appears to be a difference in the
absorption line troughs between the blue and red HiBAL samples.  The
blue HiBALs appear to have strong absorption at higher velocities
($\lambda<1510\,{\rm\AA}$) and the red HiBALs have relatively stronger
absorption closer to the emission-line peak ($\lambda\sim1520\,{\rm
\AA}$) --- similar to ``miniBALs'' or the so-called ``associated''
absorption systems \markcite{fwp+86}({Foltz} {et~al.} 1986) found in radio-loud quasars.
Although these absorption features are relatively weak, we reemphasize
that even the weak composite features must be persistent in the
majority of the input spectra, otherwise the features would be washed
out in the composites.

\subsection{Conclusions with Regard to the BALQSO Parent Population}

We suggest that the likelihood of a quasar having BAL-like intrinsic
absorption (and the type of absorption) is a function of some
intrinsic properties of the quasars (i.e., some intrinsic property or
properties determine how likely a quasar is to have BAL troughs and
how large a solid angle the flow covers).  In the optical/UV, BALQSOs
tend towards the region of multidimensional parameter space
characterized by bluer colors and larger \ion{C}{4} blueshift (which
presumably is tied to more physical properties such as luminosity
and/or accretion rate).  Although the BALQSOs do appear to favor
certain regions of parameter space, their properties span a good
portion of that space and do not appear to be disjoint from the
distribution of nonBALQSOs.  Thus we conclude that the parent samples
of BALQSOs and nonBALQSOs are the same --- at least for our optically
selected sample.

The differences between BALQSOs and nonBALQSOs (and among BALQSOs
themselves) could then be due to luminosity \markcite{cor90}({Corbin} 1990), accretion
rate \markcite{bor02}({Boroson} 2002), inclination \markcite{wmf+91}({Weymann} {et~al.} 1991), some other intrinsic
property, or a combination thereof.  It is difficult to know which of
these is primary.  The velocity asymmetry in the \ion{C}{4} emission
line is suggestive of an obscuration-induced orientation effect
(whether external, i.e. line of sight orientation, or internal,
i.e. the opening angle of the disk wind), which might be driven by
another parameter such as luminosity or accretion rate
\markcite{elv00}(e.g., {Elvis} 2000).

\section{The BALQSO Fraction \label{sec:balfractions}}

The differences in the spectra of BALQSOs compared to nonBALs must be
taken into account when determining the fraction of quasars that are
BALQSOs.  In the following sections, we first present the raw BALQSO
fraction as a function of redshift.  We then use the color properties
of BALQSOs to account for color-induced redshift selection effects
which result in a corrected BAL fraction as a function of redshift.

\subsection{Raw Fractions}

Several estimates have been made of the fraction of quasars that show
broad absorption troughs.  \markcite{wmf+91}{Weymann} {et~al.} (1991) estimated a BALQSO fraction
of 12\% (for $1.5 \lesssim z \lesssim 3.0$), while \markcite{hf03}{Hewett} \& {Foltz} (2003) found
an observed fraction of $15\pm3\%$ (and an intrinsic fraction of
$22\pm4\%$) in the same redshift range.  Our uncorrected BALQSO
fraction falls between those two results: we measure $14.0 \pm 1.0$\%
(185 BALQSOs out of 1318 quasars with $1.7 \le z \le 4.2$), similar to
the 15\% found by \markcite{kro+02}{Tolea} {et~al.} (2002) using mostly the same data set.
\markcite{sf92}{Sprayberry} \& {Foltz} (1992) estimated that the LoBAL fraction of BALQSOs is near 17\%
(1.5\% of all quasars).  Since we can identify LoBALs when \ion{Mg}{2}
or \ion{Al}{3} broad absorption is present, we can give a LoBAL
fraction for the redshift range $1.7 \le z \le 3.9$.  In this range,
our results yield an uncorrected LoBAL fraction of $13 \pm 3$\%
(24/181) among BALQSOs or about $1.9^{+0.5}_{-0.4}$\% (24/1291) of the
quasar sample.  We have not investigated the fraction of BALQSOs as a
function of radio power; see \markcite{bwg+01}{Becker} {et~al.} (2001) for such a study using the
FBQS sample.
 
\subsection{Selection Effects \label{sec:seleff}}

To determine the true BALQSO fraction, we must investigate the
completeness of the SDSS BALQSO sample and apply corrections to our
observed fraction.  For the \markcite{wmf+91}{Weymann} {et~al.} (1991) sample, it was necessary to
correct the BALQSO fraction for absorption by the BAL trough itself in
the $B$-band that was used to select quasars in the LBQS.  The SDSS
sample largely avoids that problem since SDSS quasars are selected
using an $i$-band magnitude limit.  For SDSS quasars with
$z\lesssim3.5$ no correction is needed for BAL absorption in the $i$
band.  Above that redshift a correction is needed, but instead we will
simply choose to limit our analysis to redshifts where no correction
for $i$-band absorption is needed.

Despite this selection rule, there are several other important
corrections that are needed in order to determine the ``true''
fraction of BALQSOs in our sample.

First, in terms of the initial selection of quasars, we have ignored
the fact that the limiting magnitude changes by a full magnitude
between the low-$z$ and high-$z$ ($z\gtrsim3$) selection criteria.
Also, roughly 20\% of the EDR quasars in the redshift range that can
be searched for \ion{C}{4} broad absorption were selected for
spectroscopy on the basis of ``serendipity'' \markcite{sto+02}({Stoughton} {et~al.} 2002).  Such
quasars pose a particular problem since serendipity selection uses a
fainter magnitude limit than was used for quasar selection and is
based largely on availability of fibers for a given plate, and
therefore has an extremely complicated selection function.  The raw
statistics presented here assume that the statistics of BALQSO
incidence in this set are the same as in the sample collected on the
basis of well-defined color criteria.  In the next section, we will
correct for both the redshift dependent magnitude differences and the
serendipity selection effect by imposing a more restrictive selection
criteria.

Second, reddening due to dust extinction may extinct quasars enough to
fall out of our magnitude-limited sample.  This effect could be
significant, but we make only a simple estimate of it for now (see
\S~\ref{sec:corfrac}) and defer a full analysis to a future paper.
Fortunately, the SDSS quasar selection algorithm must be more complete
to dust extincted quasars than previous optically-selected BALQSO
samples in part because of the use of the $i$-band instead of the
$B$-band as the limiting magnitude filter\footnote{The FBQS sample is
also selected using a redder filter, $E$, but the sample also includes
a blue color-cut, $O-E<2$.}.  It is also important to realize that
this effect is a function of redshift and that higher redshift BALQSOs
are more likely to fall out of the sample than lower redshift BALQSOs
since high redshift quasars sample shorter, rest-frame wavelengths,
that is, the most strongly affected by an SMC-like reddening curve.

Third, since BALQSOs appear to be redder than average, their selection
function will not depend on redshift in the same way as for the nonBAL
sample of SDSS quasars.  We show below that quasars with $2.3 \lesssim
z \lesssim 2.6$ that are very red are more likely to find themselves
embedded in the stellar locus than normal quasars and thus are less
likely to be selected as quasar candidates.  On the other hand,
quasars with $2.6 \lesssim z \lesssim 2.9$ are actually {\em more}
likely to be selected if they are very red (as long as they are not
heavily extincted, of course) since their colors will tend to push
them {\em out} of the stellar locus.  \markcite{hf03}{Hewett} \& {Foltz} (2003) have independently
commented on the importance of this effect.

In a preliminary attempt to determine corrections for color-dependent
selection effects as a function of redshift in the range $1.7 \le z
\le 3.45$, we determined theoretical colors as a function of redshift
by passing both the EDR and HiBAL composite quasar spectra through the
SDSS filter curves.  Color differences between the average quasar and
the average HiBAL quasar were then determined as a function of
redshift in the redshift ranges where both composite spectra span the
SDSS filters.  We then applied these color corrections to the
simulated quasar colors \markcite{fan99,rfs+01}({Fan} 1999; {Richards} {et~al.} 2001) that were used to test
the SDSS quasar target selection algorithm \markcite{rfn+02}({Richards} {et~al.} 2002a).  We compare
the color distribution of the simulated BALQSOs to real BALQSOs in the
bottom right-hand panel of Figure~\ref{fig:fig6}.  These simulated
BALQSO colors were processed through the SDSS quasar target selection
algorithm \markcite{rfn+02}({Richards} {et~al.} 2002a).  The results are shown in the top panel of
Figure~\ref{fig:fig10}, where we show the number of normal quasars
selected (as a function of redshift) by the solid line and the number
of ``BAL'' quasars selected by the dashed line.  Only about half of
the model ``BAL'' quasars were selected between redshifts 2.4 and 2.6,
while similarly inefficient selection of normal quasars occurs at
slightly higher redshifts, resulting in a redshift correction as shown
in the lower panel of Figure~\ref{fig:fig10}.  The estimated
correction factor rises to 2 near $z \sim 2.4$ and drops to near 0.25
at $z \sim 2.75$.  Thus the raw BALQSO fraction may be well estimated
at $z<2.2$, underestimated by a factor of 2 at $z \sim 2.4$, and
overestimated by a factor of 4 at $z \sim 2.75$.

These correction factors themselves are subject to some uncertainty.
By applying the {\em average} HiBAL reddening (curvature) correction
to each of the simulated quasars, we will obviously be over estimating
the correction for some objects and underestimating it for others;
thus we may be over- or underestimating the strength of the correction
function.  In addition, any misestimates of the correction function
may themselves be a function of redshift.  Lastly, we have made the
assumption that the amount of reddening is independent of the quasar's
intrinsic color.  If, for example, intrinsically blue or intrinsically
red quasars are more (or less) likely to be redder than average when
they are BALQSOs, then our correction fractions will be skewed.
However, if BALQSOs are really apparently redder than the average
quasar, as seems to be the case, then applying this averaged
correction is better than not applying any correction at all.

\subsection{Corrected Fractions\label{sec:corfrac}}

Using these corrections, we calculate the true BALQSO fraction as a
function of redshift by applying these corrections to the raw BALQSO
fraction as a function of redshift.  In the top panel of
Figure~\ref{fig:fig11} we show the raw fractions for both our data
set (solid line) and the \markcite{kro+02}{Tolea} {et~al.} (2002) data set (dotted line).  In
the bottom panel of the same figure, we show the BALQSO fraction as
defined by our fitted composite spectrum (FCS) method after applying
the above correction for reddening (dashed line).  The raw fraction
histogram shows a clear excess of BALQSOs for $2.7 \lesssim z \lesssim
2.9$, but this excess may simply be a result of the relative selection
probabilities for BALQSOs and nonBALQSOs --- the corrected fraction is
nearer to 10\%.  The corrected fraction at $z \sim 2.4$ approaches
50\%, but this is a single-bin spike and is likely to be an artifact
of the limited resolution of the correction function.  Note that while
we have a complete sample of BALQSOs from the EDR for $1.7 \le z \le
4.2$, we are only attempting to correct the BALQSO fractions in the
range $1.7 \le z \le 3.45$.

The BALQSO fraction as a function of redshift is also a function of
the algorithm that was used to select the quasars in the first place.
Since the EDR quasars were selected using three slightly different
algorithms \markcite{sto+02}({Stoughton} {et~al.} 2002) and since these algorithms differ slightly
from the SDSS's final quasar target selection algorithm
\markcite{rfn+02}({Richards} {et~al.} 2002a), we also correct our BALQSO fraction based on those
objects that are recovered by this final algorithm (excluding the
serendipitous targets discussed above).  There are 131 BALQSOs in the
resulting sample which is both relatively complete and relatively
homogeneous, see \markcite{rrh+02}{Reichard} {et~al.} (2003).  We have computed corrected BALQSO
fractions for the this sample of quasars; this BALQSO fraction is
plotted in the bottom panel of Figure~\ref{fig:fig11} as a thick solid
line.  Except for the spikes that are probably caused by small numbers
of objects per bin and the finite resolution of the correction
function, the BALQSO fraction is now seen to be roughly constant as a
function of redshift between $1.7 \le z \le 3.45$ and is approximately
$13.4 \pm 1.2\%$, close to our uncorrected EDR fraction ($14.0 \pm
1.0\%$) but with a considerably different redshift distribution.
Further work is needed to determine if other selection effects (or
more careful consideration of the selection effects discussed herein)
would affect the redshift dependence of the BALQSO fraction.  We also
note that making a correction based on the post-EDR selection
algorithm does not guarantee that the resulting EDR BALQSO sample is
completely homogeneous, since some quasar candidates in the EDR area
selected by the new algorithm will not (yet) have SDSS spectra.

So far, although we have corrected for dust reddening (BALQSOs being
redder than average), we have {\em not} corrected for dust extinction
(BALQSOs falling out of the magnitude limited sample), as considered
for the LBQS by \markcite{hf03}{Hewett} \& {Foltz} (2003).  A full treatment of this effect is left
for a future paper on a larger SDSS BALQSO sample.  We can, however,
estimate this correction by considering the reddest quasars.  Because
BALQSOs are typically more dust reddened, more BALQSOs than normal
quasars will have been extincted past the SDSS magnitude limit and
removed from our sample.  In \markcite{rhv+02}{Richards et al.} (2003) we estimated that those
quasars with $g-i$ colors 0.3 magnitudes redder than the median quasar
color at the same redshift comprise at least 15\% of the true quasar
population, but only 6\% of the observed quasar population.  Taking
this and the BALQSO fractions among normal and dust reddened quasars
into account, we estimate that the extinction-corrected BALQSO
fraction is 15.9$\pm$1.4\%.  This extinction correction is an
underestimate in the sense that it considers only the reddest BALQSOs.
A more refined estimate seems unlikely to equal the 22$\pm$4\% found
by \markcite{hf03}{Hewett} \& {Foltz} (2003) for the LBQS; however, given the uncertainties, the
BALQSO fractions for the SDSS and LBQS are both consistent with one of
every six quasars being a BALQSO.

Finally, in terms of understanding the origin and physics of BALQSOs,
it should be noted that in their attempt to discern between clear
outflows and the associated absorber population, it is not surprising
to find that the \markcite{wmf+91}{Weymann} {et~al.} (1991) BALQSO definition clearly
underestimates the fraction of quasars with BAL-like outflows. Thus,
even the corrected numbers above represent a lower limit to the
fraction of quasars that exhibit strong, intrinsic absorption
outflows.  The use of a revised balnicity index
\markcite{hal+02}(e.g., {Hall} {et~al.} 2002) in the future should help to alleviate this
problem.

\section{Conclusions \label{sec:conclusions}}

We find that there are substantial differences in the continuum and
emission-line properties of HiBAL and LoBAL quasars as compared to
nonBALs.  On average, HiBALs possesses an amount of reddening $\Delta
E(B-V) \approx 0.023$ relative to the average quasar, and LoBALs
possess $\Delta E(B - V) \approx 0.077$.  Wider wavelength coverage is
needed to determine the exact form of this reddening, but SMC-like
dust extinction provides an acceptable fit.  Our understanding of the
origin of these differences may be hampered to some extent by the
possibility that our BALQSO and nonBALQSO samples do not have the same
intrinsic properties.  In particular, we suggest that BALQSOs may be
intrinsically bluer (and have weaker, more strongly blue-shifted
\ion{C}{4} emission) than nonBALQSOs on the average.  The absorption
properties may also be a function of intrinsic quasar properties with
LoBALs having the most extreme properties.  Despite their differences,
we suggest that optically-selected BALQSOs are not drawn from a
different parent sample than nonBALQSOs, but rather have a bias
towards certain intrinsic quasar properties within the same parent
sample.

Using our entire sample, we find that the raw, uncorrected fraction of
HiBALs is $14\%$, and that \ion{Mg}{2} LoBALs comprise at least 1.9\%
of all quasars.  These BAL fractions are not corrected for a variety
of selection effects, most notably differential color-induced
redshift-dependent selection effects.  Using our most complete and
homogeneously selected subsample, we extend the results of
\markcite{kro+02}{Tolea} {et~al.} (2002) by attempting a correction for color-dependent
selection effects as a function of redshift.  We find that the
corrected HiBAL fraction is $13.4\pm1.2$\% (15.9$\pm$1.4\% if we
include a dust extinction correction) and is roughly constant with
redshift for $1.7 \le z \le 3.45$.  Given the constraints used to
classify BALQSOs, this fraction should be taken as a lower limit to
the fraction of quasars that have intrinsic outflows along our line of
sight.

\acknowledgements

Funding for the creation and distribution of the SDSS Archive has been
provided by the Alfred P. Sloan Foundation, the Participating
Institutions, the National Aeronautics and Space Administration, the
National Science Foundation, the U.S. Department of Energy, the
Japanese Monbukagakusho, and the Max Planck Society. The SDSS Web site
is http://www.sdss.org/.  The SDSS is managed by the Astrophysical
Research Consortium (ARC) for the Participating Institutions. The
Participating Institutions are The University of Chicago, Fermilab,
the Institute for Advanced Study, the Japan Participation Group, The
Johns Hopkins University, Los Alamos National Laboratory, the
Max-Planck-Institute for Astronomy (MPIA), the Max-Planck-Institute
for Astrophysics (MPA), New Mexico State University, University of
Pittsburg, Princeton University, the United States Naval Observatory,
and the University of Washington.  This work was partially supported
in part by National Science Foundation grants AST99-00703 and
AST03-07582 (T.~A.~R., G.~T.~R. and D.~P.~S.).  P.~B.~H. is supported
by FONDECYT grant 1010981.  This work benefited from stimulating
conversations with Julian Krolik and Zlatan Tsvetanov and from careful
review by Michael Strauss and the referee, Michael Brotherton.

\begin{appendix}

Here we demonstrate that for geometric mean composite spectra,
the composite
spectral index and reddening are the arithmetic mean spectral index
and arithmetic mean reddening of the individual spectra.  Denote $n$
individual spectra as $f_i$ ($i = 1, 2, \ldots, n$) with spectral
indices $\alpha_i$, reddenings $E_i \equiv E_{i}(B-V)$ and 
extinction law $R(\lambda) \equiv [1 + R_V] \xi(\lambda)$.  
Let all the spectra be normalized to unity at $\lambda_0$.  
The normalized spectra can then be written as
\begin{eqnarray}
f'_i (\lambda) & \equiv & \frac{f_i (\lambda)}{f_i
(\lambda_0)} = \left (\frac{\lambda}{\lambda_0}\right )^{\alpha_i}
10^{-0.4E_i [R(\lambda) - R(\lambda_0)]},
\end{eqnarray}
and thus,
\begin{eqnarray} \label{eqn:normform}
f'_i (\lambda) & \propto & \lambda^{\alpha_i} 10^{-0.4E_i[R(\lambda) - R(\lambda_0)]}.
\end{eqnarray}
The geometric mean composite $f_{gm}$ is
\begin{eqnarray}
f_{gm} (\lambda) & \equiv & \left (\prod_i f'_i (\lambda) \right
)^{1/n} \\
 & \propto & \lambda^{(\sum_i \alpha_i)/n} 10^{-0.4 (\sum_i E_i/n) (R(\lambda) - R(\lambda_0))} \\
 & \propto & \lambda^{\langle \alpha_i \rangle} 10^{-0.4 \langle E_i \rangle (R(\lambda) - R(\lambda_0))},
\end{eqnarray}
which is of the same form as Eq.~\ref{eqn:normform}.  The arithmetic
mean spectral index $\langle \alpha_i \rangle$ and reddening $\langle
E_i \rangle$ appear as the spectral index and reddening of the
geometric composite spectrum --- assuming that the form of the
reddening law is the same for all spectra.  When the reddening law is
not the same for all spectra, the geometric mean composite spectrum no
longer yields the arithmetic mean $E(B-V)$, but it still yields the
arithmetic mean absorption $\langle A(\lambda) \rangle = \langle
E(B-V) \times R(\lambda) \rangle$ at each wavelength.  That is, the
extinction factor in Equation 5 cannot be simplified beyond
$10^{-0.4<A_i(\lambda)>}$ because there is no longer a single
extinction law which can be factored out from the reddenings $E_i$,
but since $A_{\lambda} = E(B-V) \times R_{\lambda}$, for each spectrum
we have
$$ f'_i(\lambda) \propto \lambda^{\alpha_i} 10^{-0.4 A_i} $$
which yields a geometric mean of
$$ f_{gm}(\lambda) \propto \lambda^{<\alpha_i>} 10^{-0.4 <A_i>}.$$

\end{appendix}

\clearpage



\clearpage

\begin{figure}[p]
\epsscale{1.0}
\plotone{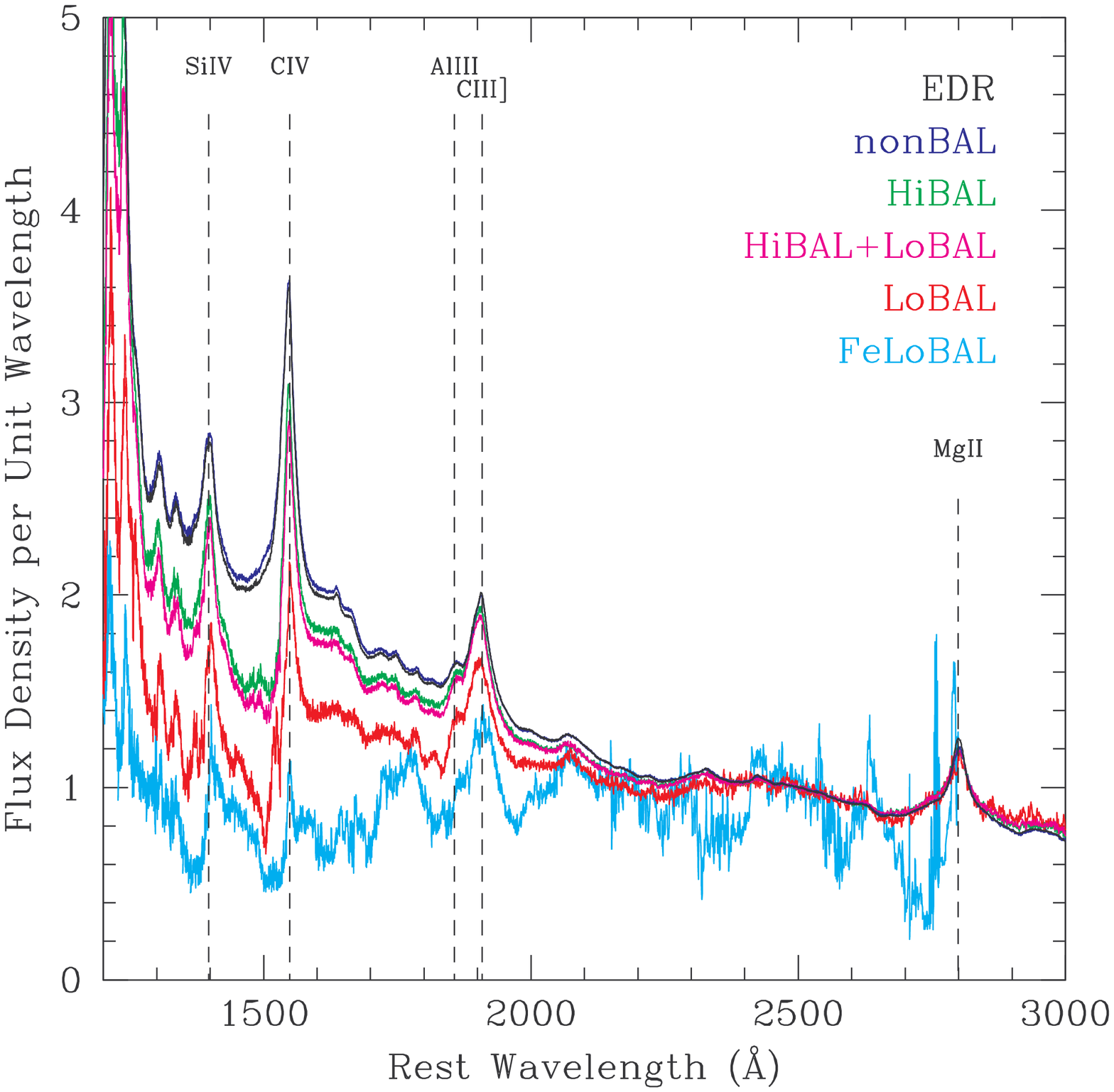}
\caption{The normalized, geometric composite spectra of --- from top
to bottom --- nonBALs (blue), the full EDR quasar sample (black),
HiBALs (excluding LoBALs and FeLoBALs, green), HiBALs and LoBALs (all
BALQSOs except for FeLoBALs, magenta), LoBALs (red), and FeLoBALs
(cyan).  The spectra are normalized at $2500\,{\rm \AA}$.  Except for
the FeLoBAL composite which exhibits excess emission at long
wavelengths, the spectra are similar at wavelengths above 2400 \AA,
but the BALQSO composite spectra show clear flux deficits at shorter
wavelengths as compared to the nonBAL composite spectrum.  The FeLoBAL
composite includes only a small number of spectra and is for
illustrative purposes only. \label{fig:fig1}}
\end{figure}

\clearpage
\begin{figure}[p]
\epsscale{1.0}
\plotone{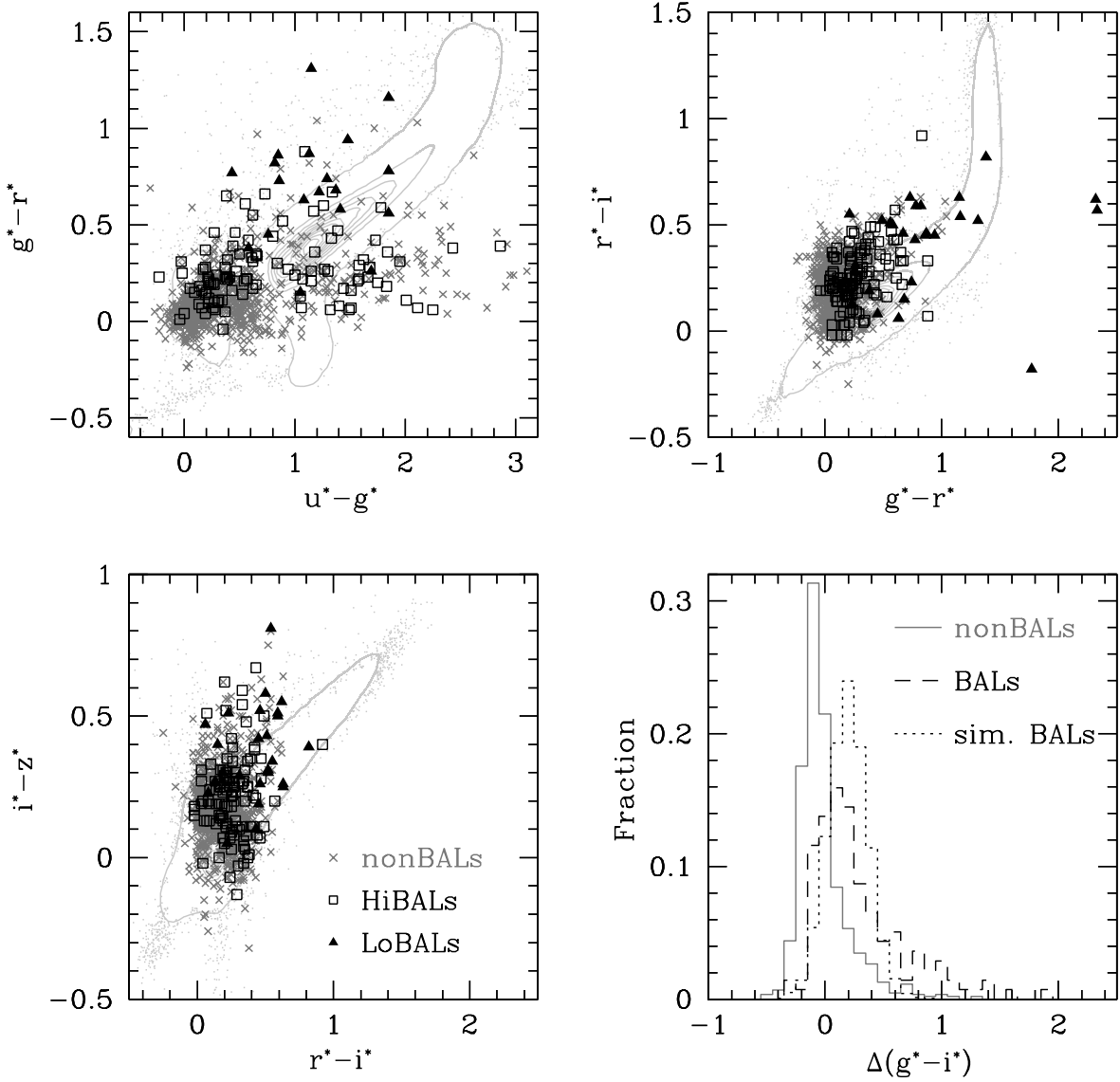}
\caption{Colors of nonBALs (dark gray ``x''s) and BALQSOs (HiBALs,
open black squares; LoBALs, filled black triangles) in SDSS
color-color-color space for quasars with $1.8 \le z \le 3.5$.  Light
gray points and contours show the distribution of all point sources in
the SDSS EDR, including quasars, stars and unresolved galaxies; only
objects with small errors in each band are included.  The lower right
hand panel shows the distribution of relative $(g^*-i^*)$ colors,
$\Delta(g^*-i^*)$, for both nonBALQSOs and BALQSOs in addition
to the distribution for the simulated BALQSOs discussed in
Section~\ref{sec:seleff}. \label{fig:fig6}}
\end{figure}

\clearpage
\begin{figure}[p]
\epsscale{1.0}
\plotone{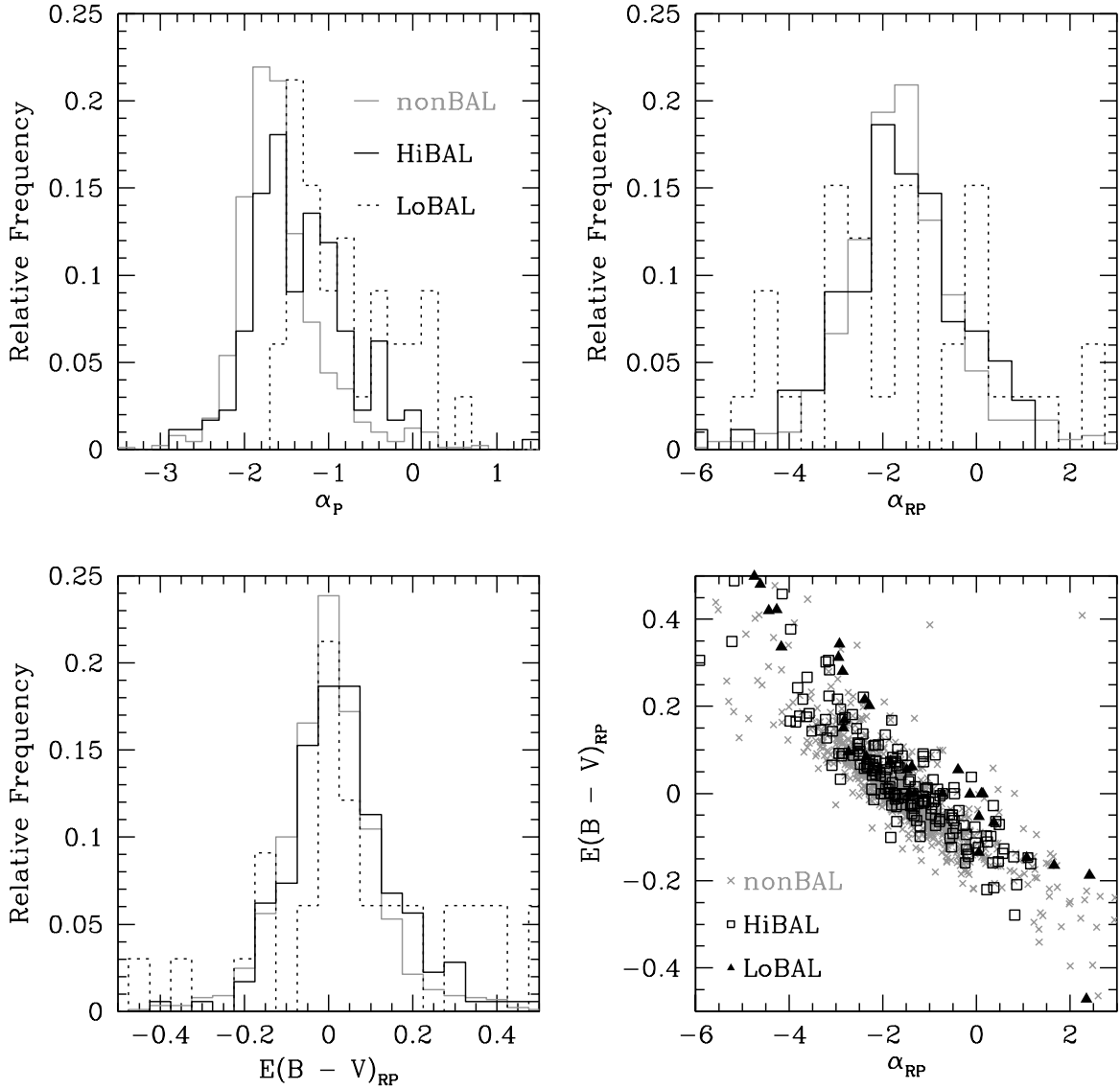}
\caption{Distribution of spectral indices and reddening values for the
$P$-fit and $RP$-fit procedures.  Solid black lines indicate HiBALs,
dashed black lines indicate LoBALs, while solid gray lines indicate
nonBALQSOs.  In the bottom right-hand panel, gray ``x'''s are for
nonBALs, open black squares for HiBALs and filled black triangles for
LoBALs. \label{fig:fig2}}
\end{figure}

\clearpage
\begin{figure}[p]
\epsscale{1.0}
\plotone{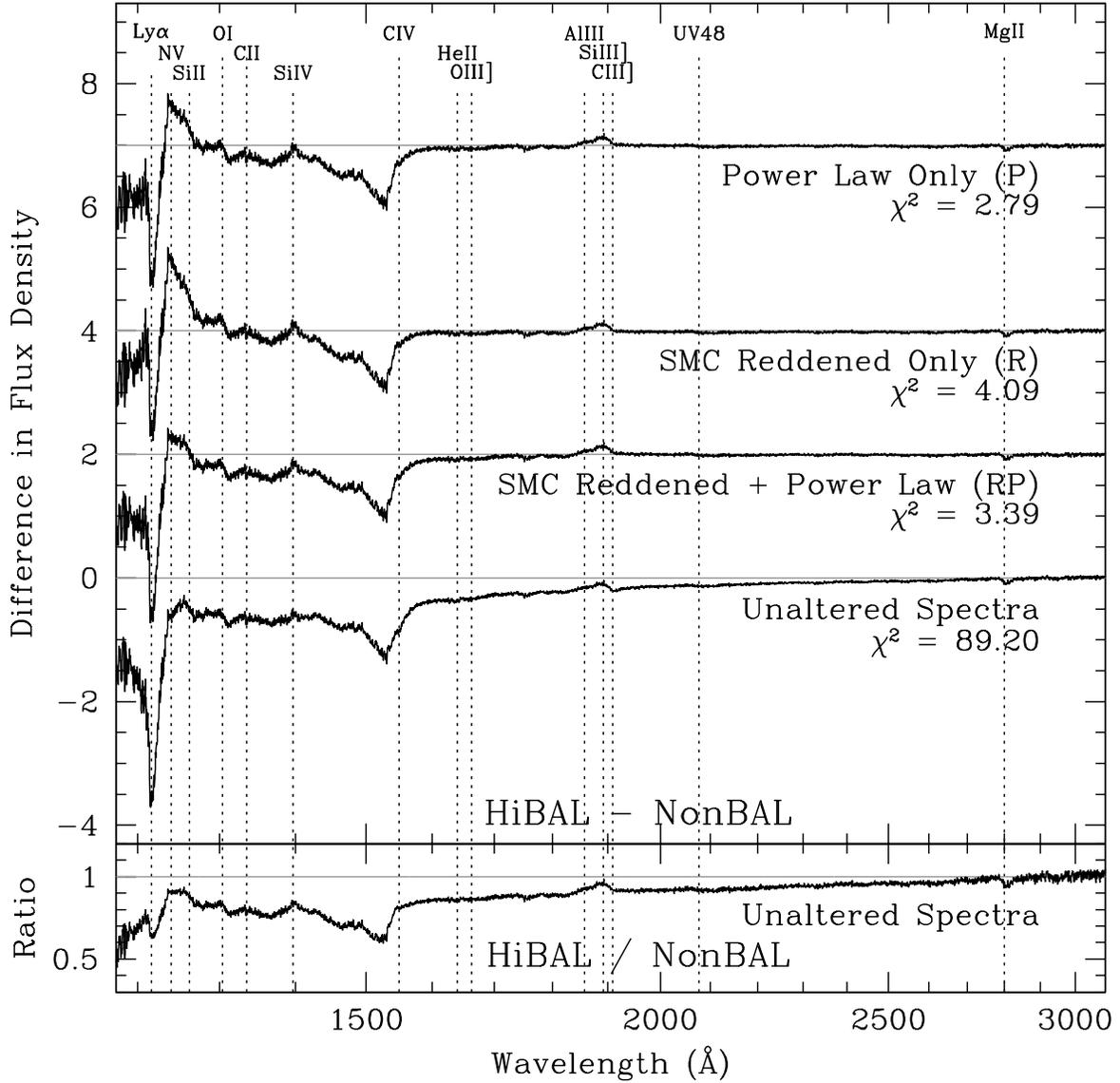}
\caption{Difference spectra for the HiBAL and nonBAL composite spectra
after matching continua via the various fitting procedures.  See text
(\S~\ref{sec:red}) for details. Compare to \markcite{wmf+91}{Weymann} {et~al.} (1991), Figures~4
and 5.  The bottom panel shows the ratio of the unaltered spectra
rather than the difference.  The $\chi^2$ values are intended as a
goodness-of-fit measure for continuum regions between 1620 and 4100
\AA.\label{fig:fig3}}
\end{figure}

\clearpage
\begin{figure}[p]
\epsscale{1.0}
\plotone{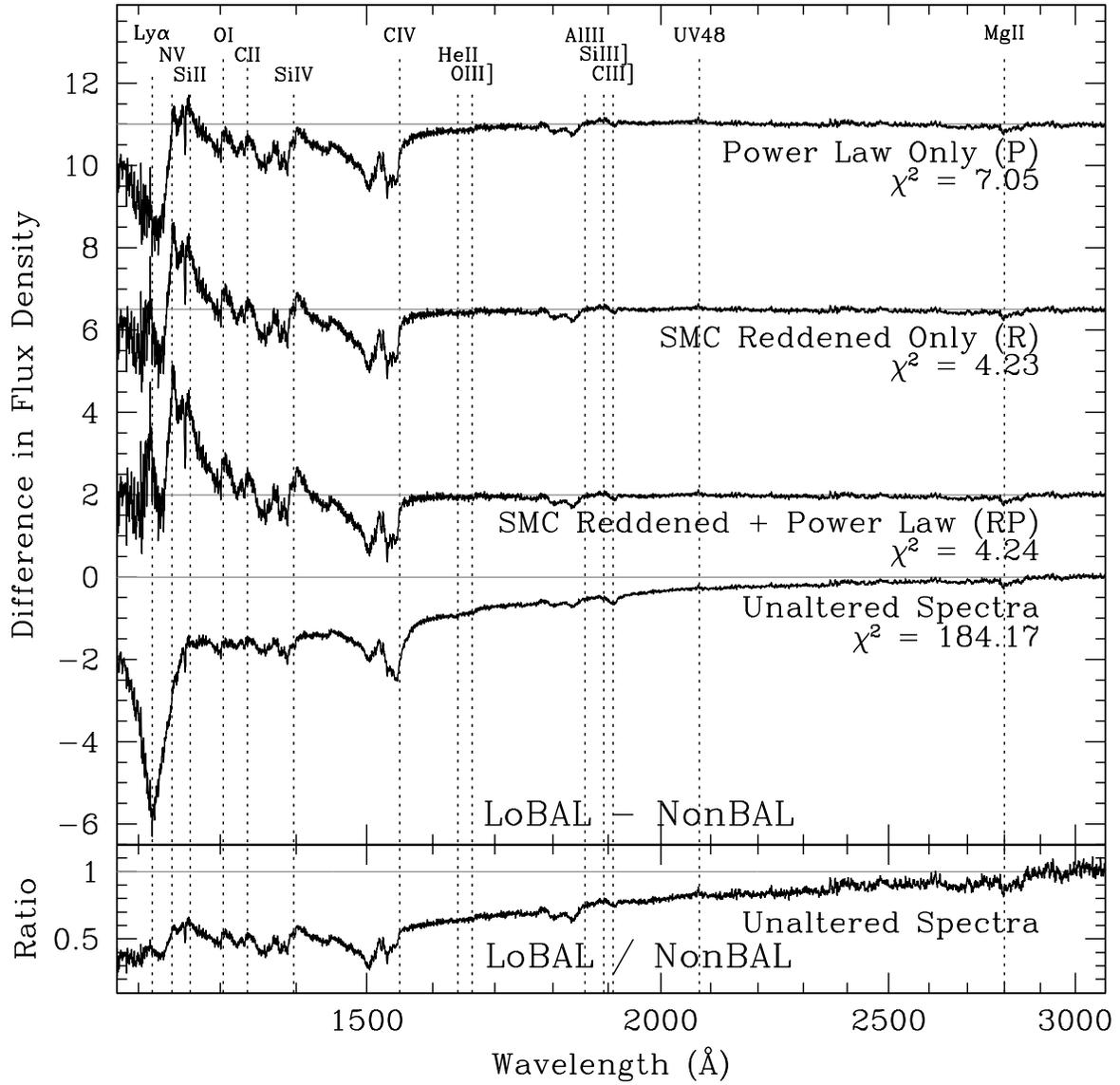}
\caption{Difference spectra for the LoBAL and nonBAL composite
spectra, as in Figure~\ref{fig:fig3}.  Compare to \markcite{wmf+91}{Weymann} {et~al.} (1991),
Figure~6.
\label{fig:fig4}}
\end{figure}

\clearpage
\begin{figure}[p]
\epsscale{1.0}
\plotone{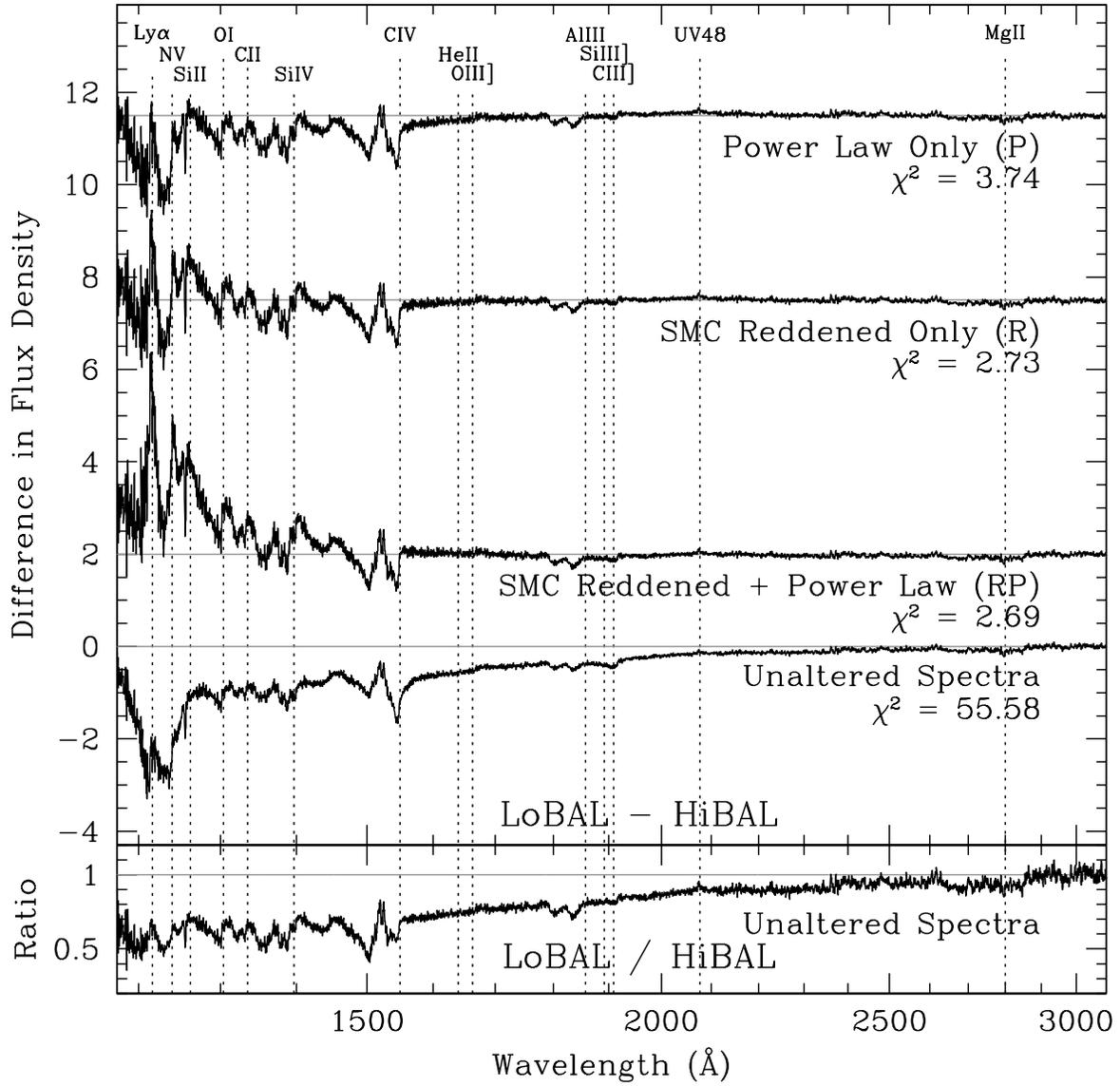}
\caption{Difference spectra for the LoBAL and HiBAL composite
spectra, as in Figure~\ref{fig:fig3}.
\label{fig:fig5}}
\end{figure}

\clearpage
\begin{figure}[p]
\epsscale{1.0} 
\plotone{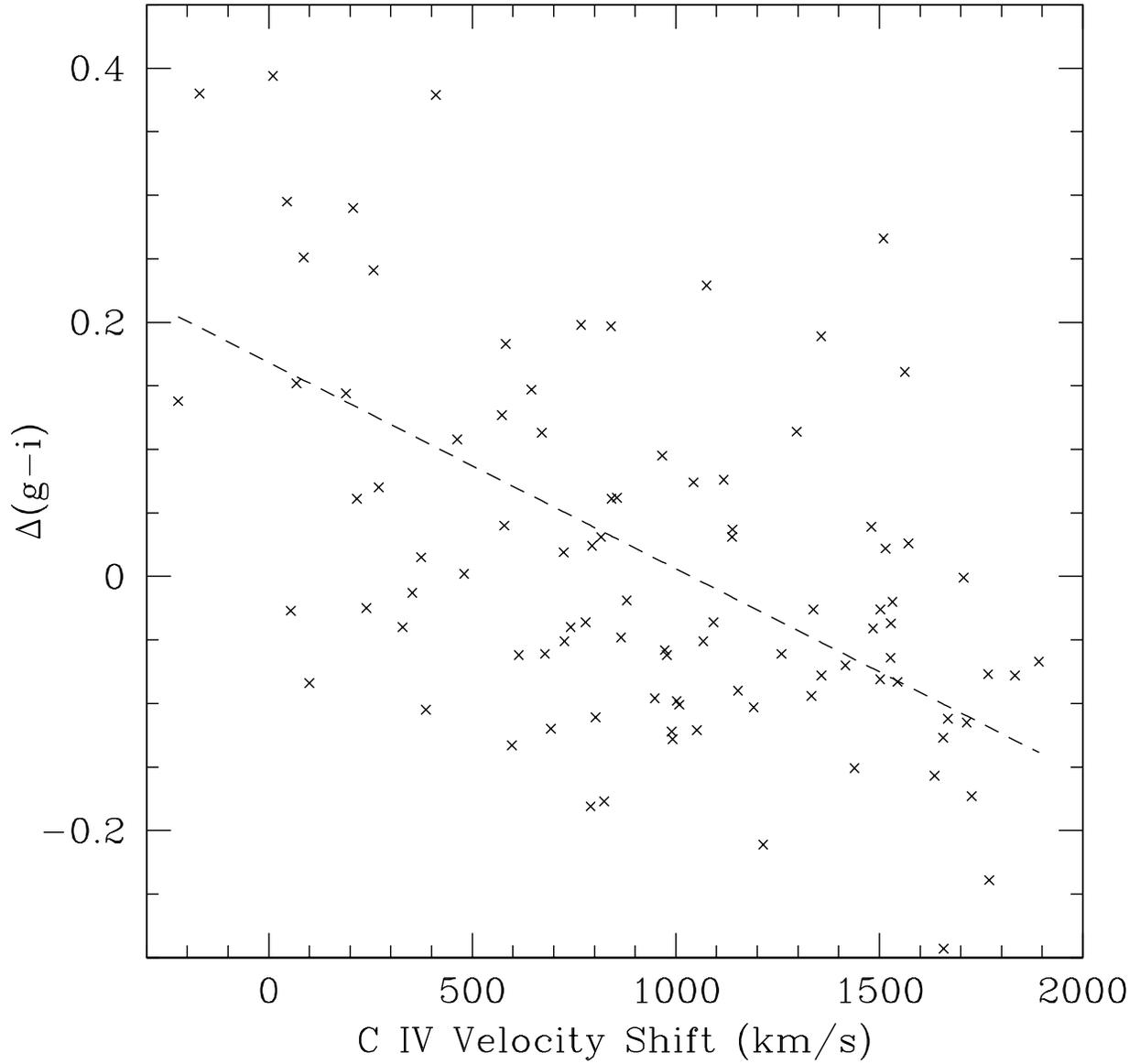}
\caption{C~IV velocity shift versus relative $g-i$ color,
$\Delta(g-i)$, for bright ($i<18.1$) EDR quasars.  Quasars with larger
C~IV emission-line blueshifts appear to have bluer intrinsic
colors. The dashed line is a least-squares fit to the
data. \label{fig:fig7}}
\end{figure}

\clearpage
\begin{figure}[p]
\epsscale{1.0} 
\plotone{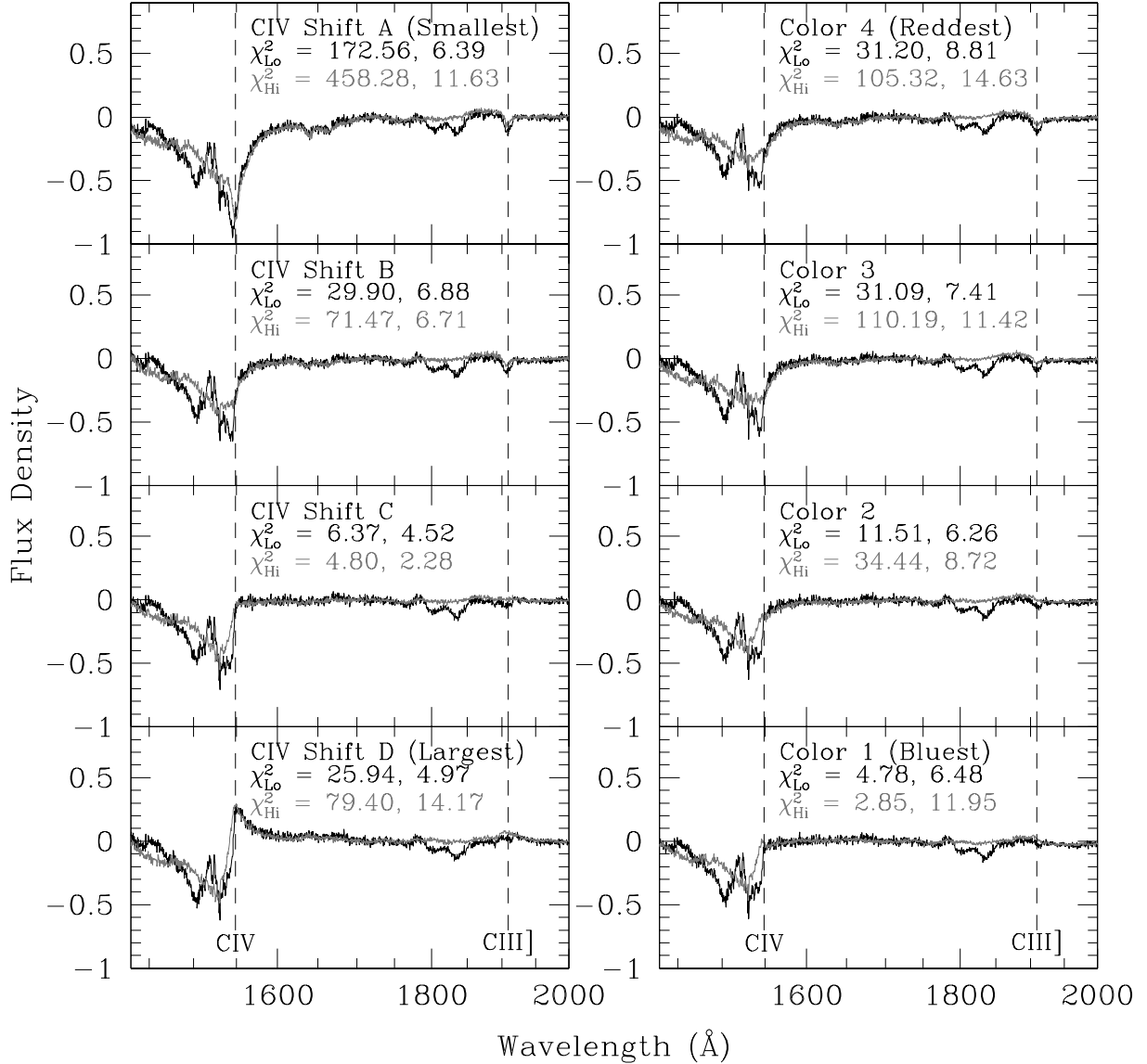}
\caption{HiBAL ({\em gray}) and LoBAL ({\em black}) composite spectra
compared with composite spectra of quasars with respect to extremes in
emission-line blueshifts and broad-band colors.  The left panels show
difference spectra with respect to \ion{C}{4} blueshift, whereas the
right panels show difference spectra with respect to color.  The two
$\chi^2$ values given in each panel are for the red wing of \ion{C}{4}
and the entire \ion{C}{3}] emission-line region, respectively.
Comparison of these two regions suggests that BALQSOs tend to be
intrinsically bluer than average and have larger than average
\ion{C}{4} emission-line blueshifts.\label{fig:fig8}}
\end{figure}

\clearpage
\begin{figure}[p]
\epsscale{1.0} 
\plotone{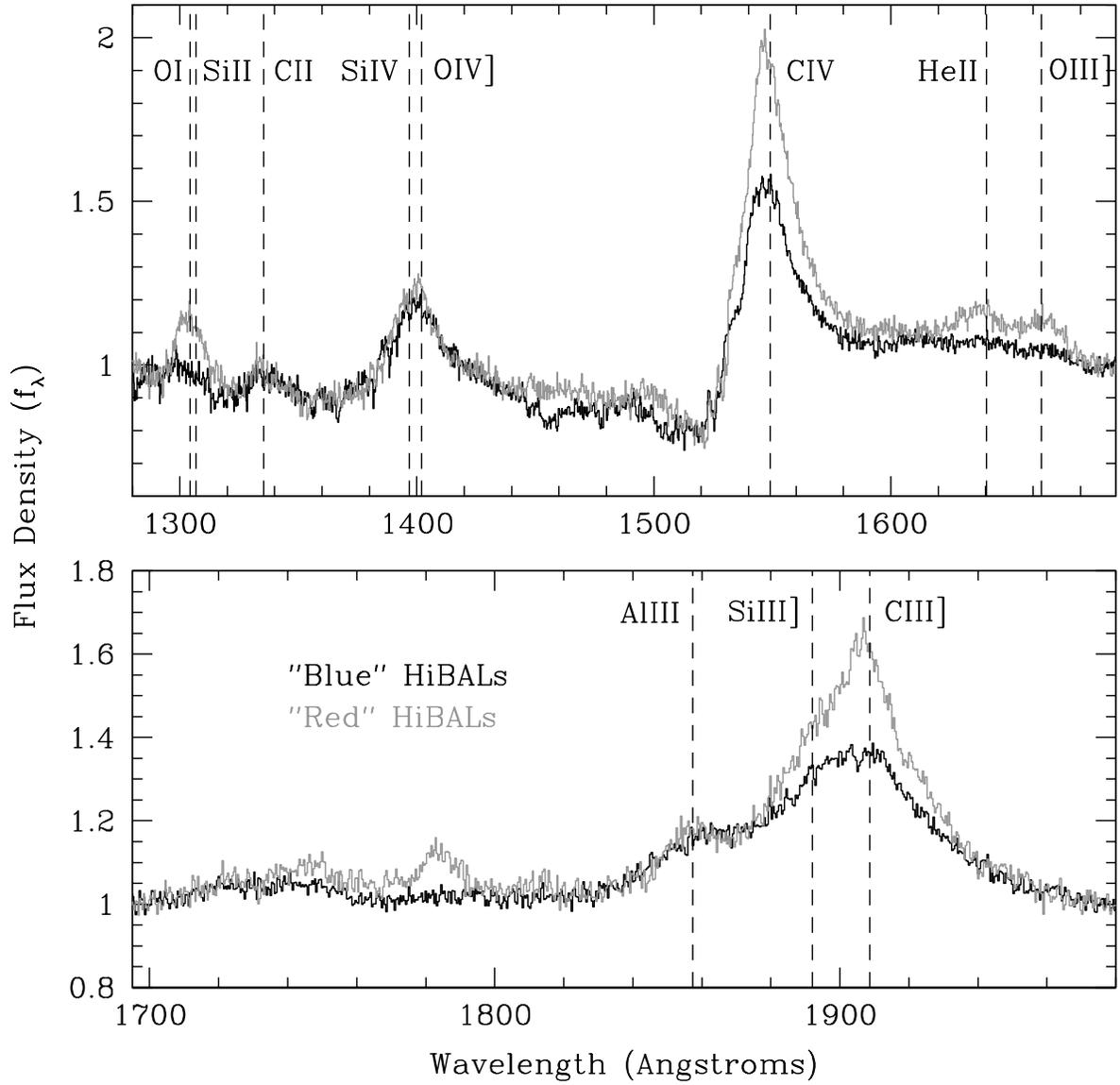}
\caption{Composite spectra of HiBALs whose $RP$-fits yield
intrinsically blue spectral indices ({\em black line}) and
intrinsically red spectral indices ({\em gray line}).  Each composite
is the average of $\sim41$ quasars. \label{fig:fig9}}
\end{figure}

\clearpage
\begin{figure}[p]
\epsscale{1.0} 
\plotone{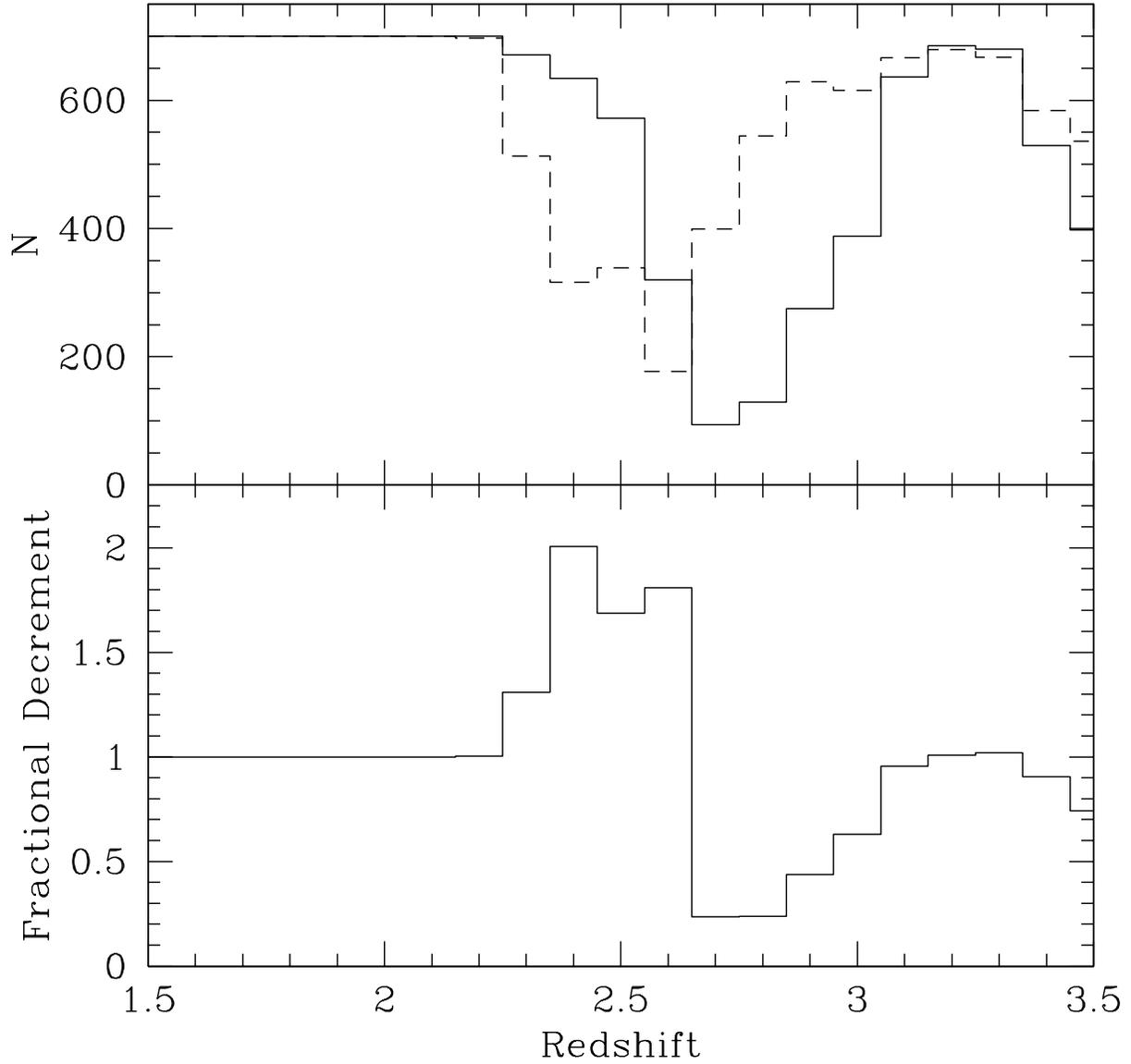}
\caption{{\em Top panel.} Number of simulated quasars selected (out of
a maximum of 700 in each redshift bin) by the SDSS quasar target
selection algorithm for normal quasars ({\em solid line}) and
simulated ``BAL'' quasars ({\em dashed line}); see text.  {\em Bottom
panel.}  The correction factor for raw BALQSO fractions due to
redshift-dependent selection effects.  \label{fig:fig10}}
\end{figure}

\clearpage
\begin{figure}[p]
\epsscale{1.0} 
\plotone{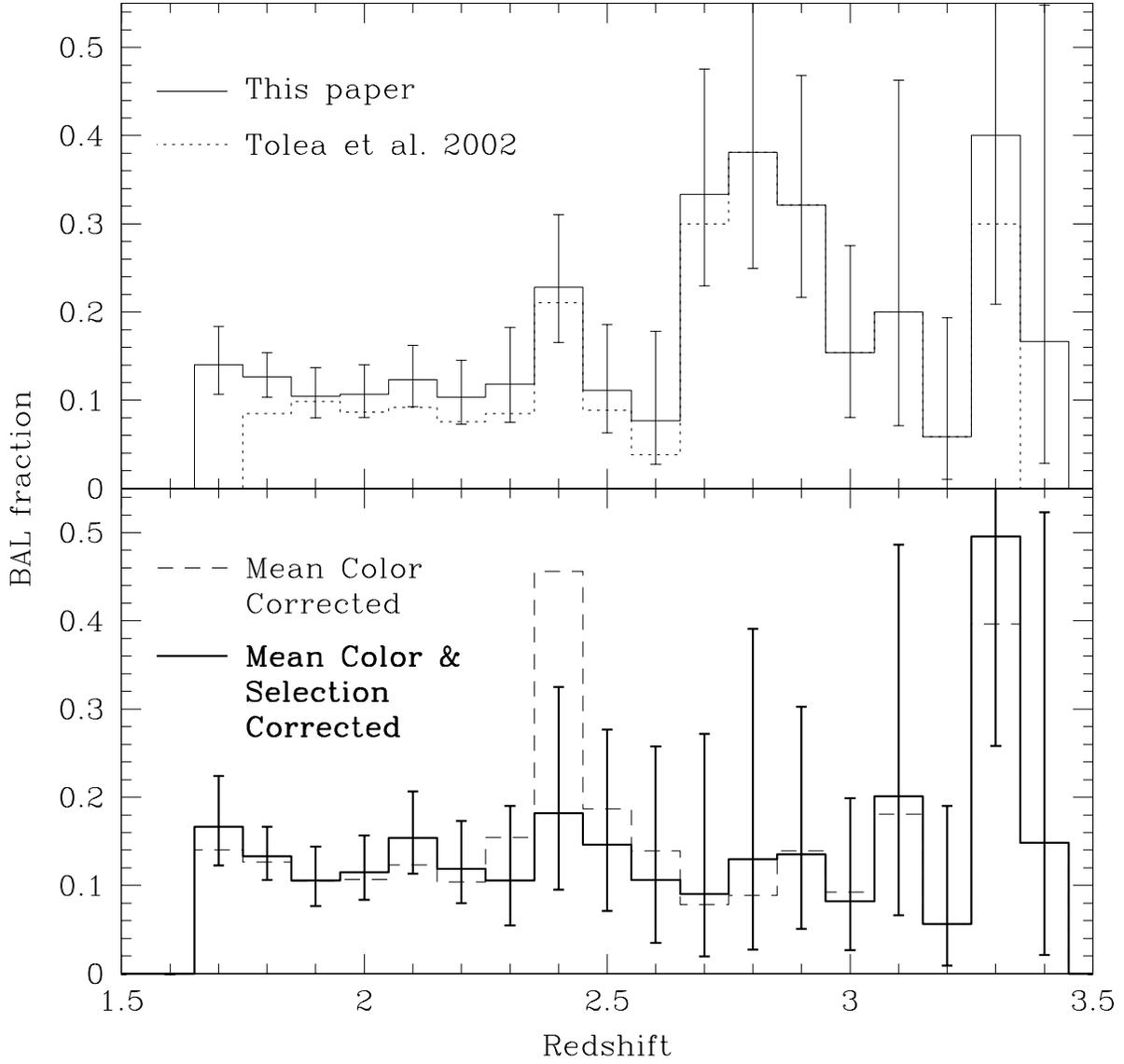}
\caption{({\em top}) The raw fraction of BALQSOs using the FCS method
of \markcite{rrh+02}{Reichard} {et~al.} (2003) ({\em solid line}) and the method of \markcite{kro+02}{Tolea} {et~al.} (2002)
({\em dashed line}).  Note that the discrepancy in the first bin
simply results from the higher minimum redshift imposed by
\markcite{kro+02}{Tolea} {et~al.} (2002) in their sample.  ({\em bottom}) The fraction of
FCS-defined BALQSOs corrected for reddening induced selection effects
({\em dashed line}); see text.  The solid line shows the distribution
of BALQSOs that were also corrected for this selection effect, but
that are restricted to those EDR quasars that were also identified as
quasar candidates by the SDSS's final quasar target selection
algorithm as described by \markcite{rfn+02}{Richards} {et~al.} (2002a).  The fully corrected
distribution yields a BALQSO fraction of $13.4\pm1.2$\% that is largely
independent of redshift.  Error bars give 68\% confidence intervals
(statistical error only).
\label{fig:fig11}}
\end{figure}

\clearpage

\begin{deluxetable}{lrrrrr}
\tabletypesize{\small}
\tablewidth{0pt}
\tablecaption{Composite Spectrum Data \label{tab:tab1}}
\tablehead{

\colhead{} &
\colhead{Number of} &
\colhead{} &
\colhead{} &
\colhead{} &
\colhead{} \\

\colhead{Class} &
\colhead{Objects} &
\colhead{$\alpha_P$} &
\colhead{$( \alpha_{RP}$} &
\colhead{$E_{RP} )$} &
\colhead{$E_R$} 

}
\startdata
EDR            & 3814 & $-1.58$ & $-1.58$ & $ 0.000$ & $ 0.000$ \\
NonBAL         &  892 & $-1.61$ & $-1.60$ & $-0.002$ & $-0.004$ \\
HiBAL          &  180 & $-1.39$ & $-1.23$ & $-0.020$ & $ 0.023$ \\
HiBAL+LoBAL    &  214 & $-1.29$ & $-1.19$ & $-0.012$ & $ 0.032$ \\
LoBAL          &   34 & $-0.93$ & $-2.01$ & $ 0.126$ & $ 0.077$ \\ 
FeLoBAL        &   10 & \nodata & \nodata & \nodata & \nodata 
\enddata 
\tablecomments{All spectral indices are in terms of $\alpha_{\lambda}$
with $f_{\lambda} \propto \lambda^{\alpha_{\lambda}}$.  $E$ represents
the reddening relative to the inherent reddening of the EDR composite
spectrum, i.e., $E \equiv E(B - V) - <E(B - V)_{EDR}>$.  The spectral
indices and reddening values are given for the three fits: $P$ for a
power law (no reddening), $RP$ for a combination power law and SMC
reddening law, and $R$ for an SMC reddening law (no change in spectral
index). See \S~\ref{sec:compred} for discussion.}
\end{deluxetable}

\end{document}